\documentclass[a4paper,11pt]{article}

\usepackage[margin=1in]{geometry} 

\usepackage{amsmath}
\usepackage{amsthm}
\usepackage{amssymb}
\usepackage{lineno}
\usepackage{subcaption}  

\usepackage{dutchcal}

\usepackage{placeins}

\usepackage{etoolbox}

\usepackage{graphicx}
\usepackage{bm}
\usepackage{lineno}
\usepackage{caption}

\usepackage{booktabs}

\usepackage[breaklinks]{hyperref}

\usepackage{multirow}

\usepackage[linesnumbered,ruled,vlined]{algorithm2e}

\usepackage{comment}
\usepackage{xargs}                      

\hypersetup{
	unicode,
	colorlinks=true,
	linkcolor=blue,
	filecolor=magenta,      
	citecolor=blue,
	urlcolor=cyan,
	pdfauthor={ M. A. Mendez},
	pdftitle={Linear and Nonlinear Dimensionality Reduction from Fluid Mechanics to Machine Learning},
	pdfsubject={Linear and Nonlinear Dimensionality Reduction from Fluid Mechanics to Machine Learning},
	pdfkeywords={article, template, simple},
	pdfproducer={LaTeX},
	pdfcreator={pdflatex}
	colorlinks=true,
	citecolor=blue,
	linkcolor=blue,
	filecolor=magenta,      
	urlcolor=blue,
}

\usepackage[most]{tcolorbox}
\usepackage{booktabs}
\usepackage{amsmath}
\usepackage{wrapfig}

\usepackage{mathtools}
\tcbset{enhanced,colback=cyan!2!white,colframe=cyan!90!white,coltitle=blue!20!black,fonttitle=\bfseries}

\usepackage{hyperref}
\hypersetup{
	colorlinks=true,
	linkcolor=blue,
	filecolor=magenta,      
	citecolor=blue,
	urlcolor=cyan,
}

\usepackage{mathrsfs} 

\usepackage{siunitx}
\newcolumntype{Z}[1]{S[scientific-notation=fixed, fixed-exponent=#1, table-format=1.2]}

\title{\LARGE \textbf{Characterisation and extension of a rigid body dynamics solver coupled with OpenFOAM for flight performance analysis of flapping-wing drones}}

\author{
  Romain Poletti$^{1,2}$\thanks{Corresponding author: romain.poletti@vki.ac.be} \and
  Emanuele Bombardi$^{3}$ \and
  Lilla Koloszar$^{1}$ \and
  Miguel Alfonso Mendez$^{1,3,4}$\and
  Joris Degroote$^{2}$ 
}

\date{%
\begingroup
\small
\textit{%
$^1$ Environmental and Applied Fluid Dynamics, von Karman Institute for Fluid Dynamics, Belgium \\
$^2$ Department of Electromechanical, Systems and Metal Engineering, Ghent University, Belgium \\
$^3$ Aero-Thermo-Mechanics Laboratory, École Polytechnique de Bruxelles, Université Libre de Bruxelles, Belgium} \\
$^4$ Experimental Aerodynamics and Propulsion Lab, Universidad Carlos III de Madrid, Spain \\
\endgroup
}

\begin{document}
\maketitle
	
\begin{abstract}

The extraordinary aerial agility of hummingbirds and insects continues to inspire the design of flapping-wing drones. To replicate and analyze such flight, computational fluid dynamics (CFD) simulations that couple flow solvers with rigid body dynamics are essential. While OpenFOAM offers tools for these multiphysics simulations, two key limitations remain: (1) a lack of thorough verification and performance characterization, and (2) the reliance on torque-based control for wing motion, which is impractical for parametric studies and real-time control. The developments are tested with a four and a five degrees of freedom flapping-wing drone equipped with a rigid, semi-elliptical wing. Ascending flight motions are simulated using the overset method, a moving background grid, and an LES model. Parametric studies demonstrate the independence of the grid and integration schemes, while profiling analyses identify the overset method as the computational bottleneck. The drone trajectories are compared with those from a literature quasi-steady model, and the body-wing interaction is analyzed in detail.

\vspace{7mm}
\end{abstract}

\section{Introduction}

Hummingbirds and insects exhibit exceptional flight maneuverability, seamlessly transitioning from stable hovering in wind gusts to rapid escape maneuvers, including upside-down flight \cite{Cheng2016}.
Their remarkable flights have inspired the design of flapping-wing drones, offering significant potential for rescue operations in tight and hazardous spaces. However, a significant performance gap remains between engineered fliers and their biological counterparts. This gap stems from the incomplete characterization of the low-Reynolds, unsteady flow generated by the flapping wings coupled with the drone motion. Bridging this gap is crucial for our research, which aims to optimize the wing kinematics of hummingbird-size drones in ascending and hovering flights. High-fidelity environments are then required to simulate the unsteady aerodynamic phenomena during the drone flights and refine the optimizer predictions.

Simulating FWMAV flight requires coupling the drone’s equations of motion with an aerodynamic model. Semi-empirical, quasi-steady formulations are commonly used for this purpose, as they enable efficient force estimation suitable for stability and control analyses \cite{Taylor2002,Karasek2012,Schenato2003,Deng2006}. However, these models are not designed to capture unsteady aerodynamic phenomena associated with dynamic wing motions—such as transient leading-edge vortices \cite{Chen2020}, rotational circulation \cite{Meng2015}, added mass effects \cite{Liu2020}, and wing-wake interactions \cite{Lee2018}—or to resolve detailed flow structures and body–wing coupling effects.

CFD simulations have been widely used to investigate unsteady mechanisms in flying animals such as hummingbirds \cite{Song2014}, mosquitoes \cite{Aono2008}, and hawkmoths in \cite{Liu2021}.
These simulations typically rely on dynamic meshing techniques that adapt the computational grid to the wing's motion. Among them, the overset method \cite{Bomphrey2017,Cai2021}, the immersed boundary method \cite{Luo2012,Zheng2013} and the Arbitrary Lagrangian Eulerian methods \cite{Bos2013_2,Deng2016} have been applied to simulate tethered and hovering flights. However, free flight simulations remain relatively scarce, likely due at least in part to the high computational cost associated with the need to resolve both complex kinematics and flow features in strongly unsteady regimes.

These simulations couple a flow solver and a rigid body dynamics (RBD) solver using a partitioned approach. Within each time step, the Navier-Stokes equations are solved and the pressure is integrated over the drone's surface. The resulting loads serve as boundary conditions for the drone's dynamic system, which is subsequently solved to compute the drone's motion. Although various coupling schemes exist \cite{Delaisse2021}, weak coupling strategies have been predominantly used \cite{Yeo2010,Zhang2019}. This hinders their applications to highly dynamics flights with strong added mass effects.

For low-dynamic flights, the flow equations are often solved around isolated wings and the resulting aerodynamic forces are applied to the body's equations of motion \cite{Aono2008,Wu2009,Zhang2019}.
While some studies do model the flow around both the drone body and its wings \cite{Liu2010,wu2014,Yao2018,Yao2019v2}, they often neglect inertial coupling by ignoring the wings' mass and inertia \cite{Taha2012,Orlowski2011}.
The equations of motion are then those of a standard aircraft with external aerodynamic forces applied by the wings \cite{Etkin1995}. 
Given the large size of the flapping-wing drone investigated in this research, this assumption may not hold \cite{Taha2012,Orlowski2011}. The most complete approach is to consider the multibody equations of the drone, solving simultaneously for the motion of both the drone body and its wings. Only a few works have coupled this dynamic model with flow solvers using commercial software \cite{Vanella2010,Bakhshaei2021} or in-house developments \cite{Xue2023,Biswal2019}.

Consequently, these contributions offer limited benefit to the open-source community. In many cases, the underlying algorithms are insufficiently documented, making it difficult to assess their applicability or limitations. In this work, we address this gap by developing and thoroughly documenting a high-fidelity simulation environment within OpenFOAM that integrates fluid and rigid body dynamics solvers.

OpenFOAM provides two rigid body dynamic solvers within the \textit{sixDofRigidBodyMotion} or the \textit{rigidBodyDynamics} libraries. 
The six degrees of freedom library solves the Newton–Euler equations for a single rigid body subjected to external forces and constraints, including fluid dynamic loads, and has been applied in recent studies to analyze the motion of floating structures \cite{Chen2024, Karola2024}.
The \textit{rigidBodyDynamics} library generalises the six degrees of freedom library. It implements the articulated-body algorithm (ABA) \cite{Siciliano2008}, enabling the simulation of arbitrary motion for a multibody system defined by a set of bodies connected by a set of degrees of freedom. However, the current ABA implementation is limited to multibody systems with passive joints, i.e. joints driven by the system dynamics and external loads. Therefore, it cannot impose prescribed kinematics on the drone wings while simultaneously solving for the body motion. 

This article extends the capabilities of the ABA algorithm by implementing active joints for which the user can impose their kinematics as inspired by \cite{Docquier2013,Hu2005,Featherstone2014}. 
The proposed extended ABA (eABA) solves (1) a forward dynamics problem to compute the motion of the drone body and (2) an inverse dynamics problem to actuate the wing kinematics according to user-defined positions, velocities, and accelerations.
The algorithm is included in a high-fidelity environment that combines the overset method, a moving background grid, and a Large-Eddy Simulation (LES) model to simulate the flow around a drone equipped with rigid, semi-elliptical wings. 

This article is organized as follows. Section 2 details the \textit{rigidBodyDynamics} library's approach to formulating and solving the equations of motion. Section 3 describes the new implementations in the \textit{rigidBodyDynamics} library to actuate a subset of the degrees of freedom from a multibody system. Section 4 presents two test cases: a single-wing drone and a body-wing drone. The first test case is used in section 5 to verify the extended articulated-body algorithm and to perform a few parametric studies. Section 6 explores the dynamics of the open-loop, vertical ascending flight of the body-wing drone. Section 7 discusses the main contributions of this work and suggests further developments.

\section{Theoretical background of the rigid body dynamics library}
Three main solvers make up the CFD environment that computes the motion of a multibody system subjected to an external flow: (1) a rigid body dynamics solver to solve the equations of motion and predict the body motion, (2) a dynamic mesh solver to adapt the grid according to the body state and (3) a flow solver to compute the forces acting on the bodies. 
\textcolor{black}{
Figure \ref{fig:diagram} shows these three key components for the original implementation of the \textit{overPimpleDyMFoam} solver within OpenFOAM v2206.
This section describes the governing equations of the rigid body dynamics solver, linking them to the \textit{rigidBodyDynamics} library available in OpenFOAM.}
Blocks (2) and (3) are briefly described at the end of the section and the reader is referred to \cite{Hadzic2006,Ferziger2002} for more details.


\subsection{System definition}
\begin{figure}[!h]
\centering
\includegraphics[width=0.8\textwidth]{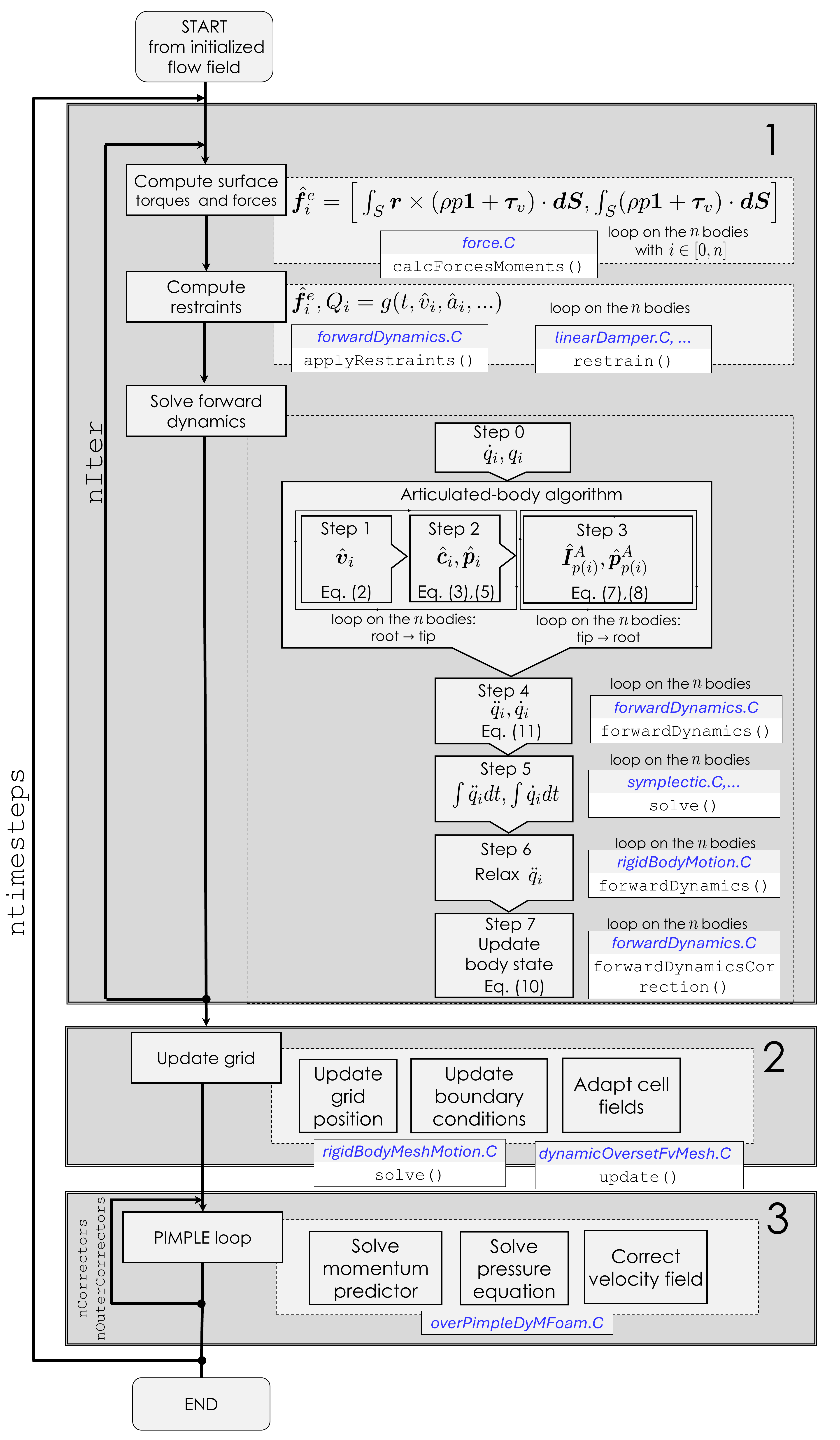}
\caption{\textcolor{black}{Diagram representing the main operations performed by (1) the rigid body dynamic solver relying on the ABA algorithm, (2) the dynamic mesh method relying on the overset method, and (3) the flow solver using pimple loops.}}
\label{fig:diagram}
\end{figure}

\begin{figure}[!h]
\centering
\includegraphics[width=0.6\textwidth]{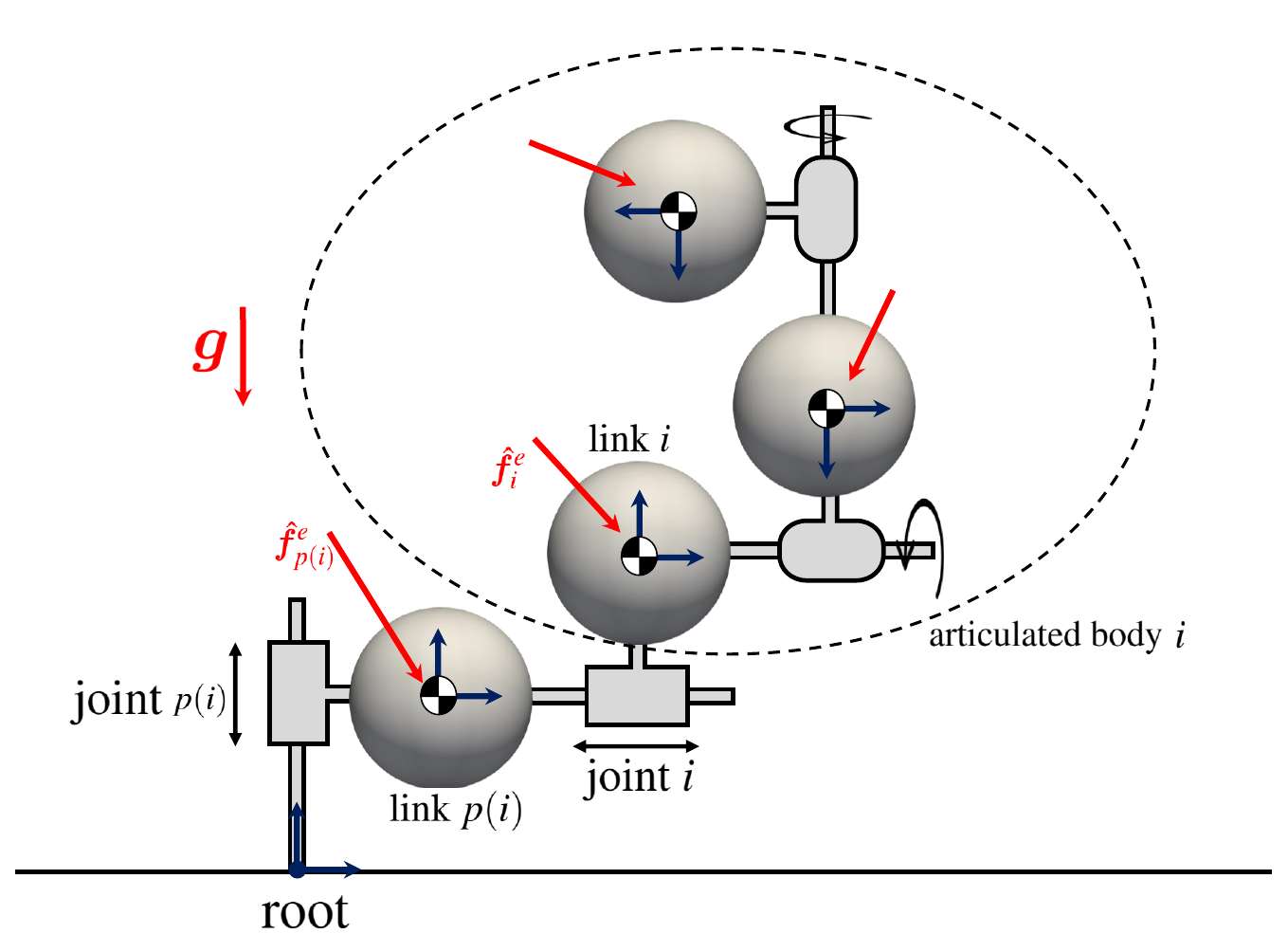}
\caption{Schematic of the kinematic tree of a multibody system with four links and four joints, illustrating the articulated body concept.}
\label{fig:scMBS}
\end{figure}

Before solving the dynamics of the multibody system, that is before the solver enters the first loop in Figure \ref{fig:diagram}, the \textit{rigidBodyDynamics} library is used to build a multibody system with entries from the \textit{dynamicMeshDict}. 
Figure \ref{fig:scMBS} illustrates one such multibody system. It is defined as a set of $n$ bodies (also called links and represented as spheres) connected through $n-1$ joints (also called articulations and represented by rectangles or cylinders \cite{Flores2015}) that allow the relative motion between the bodies.The bodies are slightly offset from the joints with a grey bar for clarity purposes. 
Each body $i$ is connected with one joint to its parent body $p(i)$ and this article considers only tree-like system, without kinematic loops formed by bodies.

The first body ($i=0$) is always a (static) \textit{root} body followed by bodies predefined in the \textit{rigidBodyDynamics} library (\textit{sphere}, \textit{cuboid}, etc). This article focuses on the most general body type, the \textit{rigidbody} that has a point mass $m$, a center of mass position $\bm{x}_c$, an inertia matrix $\bm{I}$ (defined with respect to the origin) and a frame of reference defined by its joint. 

The joints are either revolute or prismatic, and have a single degree of freedom along their axis of revolution or translation, respectively. Multiple joints can be chained between bodies to form a \textit{composite} joint. In this case, mass-less bodies are inserted in between successive joints to conform with the Newton-Euler formalism of the equations of motion (see next section). The reader is referred to \cite{rbd} for the complete functionalities of a multibody system in OpenFOAM. 





\subsection{Forward dynamics}

Based on the initial states of the multibody system and the forces applied to it, the joint accelerations are computed by solving the equations of motion. 
The main steps for this computation, known as the forward dynamics, are illustrated in the first block of Figure \ref{fig:diagram} and rely on the articulated-body algorithm. The reader is referred to \cite{Mirtich1996,Featherstone2014} for a comprehensive overview of the mathematical framework. Here, we briefly recall the seven key steps to provide foundations for the implementations discussed in section \ref{sec:impl}.


\subsubsection*{Step 0: Computation of the external and internal forces and torques} 
At the beginning of each time step, a resultant torque and force vector is computed for each link $i$ based on the pressure and shear stress that the flow applies on its surface. Those torques and forces are called \textbf{external} and they are gathered into a \textbf{spatial} vector $\hat{\bm{f}}^e_i \in \mathbb{R}^{6\times1}$ that stacks their three components:

\begin{equation}\label{eq_fe}
\hat{\bm{f}}^e_{i} = \biggl [\int_{{S}} \bm{r} \times  (\rho p \bm{1} + \bm{\tau}_\nu) \cdot d\bm{S},\int_S(\rho p \bm{1} + \bm{\tau}_\nu) \cdot d\bm{S} \biggr],
\end{equation} where the hat notation indicates a spatial vector using Plücker coordinates and bases. $p$ is the kinematic pressure, $\bm{1}$ is the identity tensor, $\bm{\tau}_\nu$ is the shear stress tensor (including fluid density multiplication), $\rho$ is the \textcolor{black}{fluid} density, $\bm{r}$ is the centre position of the body patch faces taken with respect to the origin, $d\bm{S}$ is the area of the body patch faces multiplied by its unitary normal vector, and $S$ is the body surface. The forces are applied to the body's center of gravity and defined in the inertial frame (static frame).\\
The spatial external force may be complemented by user-defined restraint forces $\hat{\bm{f}}^p_{i}$.
For example, the user can attach a spring or a damper to a body to generate additional external forces.  
Similarly, the restraint library can be used to apply an \textbf{internal} torque, $Q$, along the joint axis. For example, the user could model articulations with friction.

\subsubsection*{Step 1: Computation of the link velocities} The ABA algorithm's first step aims to compute each link's velocity from the multibody system.  The velocity of each link $\hat{\bm{v}}_i$ is defined by the velocity of the parent link $\hat{\bm{v}}_{p(i)}$ and the joint velocity $\dot{q}_i$, which models the relative velocity between the two bodies: 
\begin{equation}\label{eq_vi}
    \hat{\bm{v}}_i  = {}_i\hat{\bm{X}}_{p(i)} \hat{\bm{v}}_{p(i)} + \hat{\bm{\Phi}}_{i} \dot{q}_i, 
\end{equation} where $\hat{\bm{v}}_i$ is a spatial vector with angular velocities in the first three entries and linear velocities in the last three. The velocities are expressed in the link $i$ frame and ${}_i\hat{\bm{X}}_{p(i)}\in\mathbb{R}^{6\times6}$ is a spatial transformation matrix from the parent to the current link frame, $\hat{\bm{\Phi}}_{i}\in \mathbb{R}^{6\times1}$ is a spatial vector that converts the joint velocity $\dot{q}\in \mathbb{R}$ into an angular and linear velocity for the link. 

Equation \eqref{eq_vi} defines the first loop of the ABA algorithm: each link velocity is recursively defined spanning the multibody system from the root towards the tip body.


\subsubsection*{Step 2: Computation of the velocity-product acceleration, isolated inertia, and isolated bias forces}
Step 2 involves pre-computing three link variables used in step 3. Firstly, the velocity-product acceleration is defined based on the joint and link velocities (step 1):

\begin{equation} \label{eq_c}
\hat{\bm{c}}_{i} = \hat{\bm{v}}_{i} \times \hat{\bm{\Phi}}_{i} \dot{q}_i\,.
\end{equation}
Secondly, the (constant) spatial isolated inertia matrix $\hat{\bm{I}}_i$ is defined with the mass and inertia matrix of the link as:

\begin{equation}\label{eq_Ii}
\hat{\bm{I}}_i =
\begin{bmatrix}
\mathbf{I}^{c}_i + m_i \bm{S}^T(\bm{x}_c)\bm{S}(\bm{x}_c) & m_i \bm{S}(\bm{x}_c) \\
m_i \mathbf{S}^T(\bm{x}_c) & m_i \mathbf{1}
\end{bmatrix},
\end{equation}
where $m_i$ is the link mass, $\mathbf{I}^{c}_i$ is the moment of inertia tensor about the center of mass and $\bm{S}(\bm{x}_c)$ is the skew matrix for the centre of mass position $\bm{x}_{c}$.

Thirdly, the isolated bias force gathers all the terms that are not a function of the link acceleration in the equation of motion:

\begin{equation}\label{eq_pI}
    \hat{\bm{p}}_{i} = \hat{\bm{v}}_{i} \times \hat{\bm{I}}_{i} \hat{\bm{v}}_{i}  - {}_i\hat{\bm{X}}^{-T}_{0} \hat{\bm{f}}^e_{i}.
\end{equation}
The first term is the inertia term due to the frame rotation, and the second term is the spatial vector of external forces (step 0) transformed in the link frame with the spatial matrix ${}_i\hat{\bm{X}}^{-T}_{0}$, with $^{-T}$ denoting the inverse of the transpose of a matrix. 

This second step is performed during the first loop defined in step 1. 

\subsubsection*{Step 3: Computation of the articulated inertia and bias forces}

The Newton-Euler equations for each parent link $p(i)$ formulated at the center of mass can be written as (see\cite{Featherstone2014}):

\begin{equation}\label{eq_fpi}
    \hat{\bm{f}}_{p(i)} =  \hat{\bm{I}}^{A}_{p(i)} \hat{\bm{a}}_{p(i)} + \hat{\bm{p}}^{A}_{p(i)}, 
\end{equation} where $\hat{\bm{f}}_{p(i)}$ is the spatial force applied at the joint and transformed at the center of mass of a parent body. 
$\hat{\bm{I}}^{A}_{p(i)}$ and $\hat{\bm{p}}^{A}_{p(i)}$ are the inertia and bias forces of the articulated body which is the subtree that contains all the links from the body $i$ to the tip body $n$ (Figure \ref{fig:scMBS}):

\begin{equation}\label{eq_IA}
\hat{\bm{I}}^{A}_{p(i)} = \hat{\bm{I}}_{p(i)} +   {}_{p(i)}\hat{\bm{X}}_{i}\Bigg( \hat{\bm{I}}^{A}_i - \frac{\hat{\bm{I}}^{A}_i \hat{\bm{\Phi}}_{i} \hat{\bm{\Phi}}^T_{i} \hat{\bm{I}}^{A}_i }{\hat{\bm{\Phi}}^T_{i}  \hat{\bm{I}}^{A}_i \hat{\bm{\Phi}}_{i}}\Bigg){}_{i}\hat{\bm{X}}_{p(i)},
\end{equation}

\begin{equation}\label{eq_pA}
\hat{\bm{p}}^A_{p(i)} = \hat{\bm{p}}_{p(i)} +   {}_{p(i)}\hat{\bm{X}}_{i}\Bigg( \hat{\bm{p}}^{A}_i +  \hat{\bm{I}}^{A}_i \hat{\bm{c}}_i + \frac{\hat{\bm{I}}^{A}_i \hat{\bm{\Phi}}_{i} \Big[Q_i - \hat{\bm{\Phi}}^T_{i}(\hat{\bm{p}}^{A}_i + \hat{\bm{I}}^{A}_i \hat{\bm{c}}_i ) \Big] }{\hat{\bm{\Phi}}^T_{i}  \hat{\bm{I}}^{A}_i \hat{\bm{\Phi}}_{i}}\Bigg). 
\end{equation}The articulated bias force depends on $Q_i$, the magnitude of the spatial joint force defined as:

\begin{equation}\label{eq_Qi}
    Q_i = \hat{\bm{\Phi}}^T_i \hat{\bm{f}}_i.
\end{equation}
For an unconstrained joint, $Q_i$ is zero so that the joint velocity $\dot{q}_i$ is only defined by the multibody system dynamics. The joint force can also be user-imposed in step 0.

Equations \eqref{eq_IA} and \eqref{eq_pA} are defined recursively based on the inertia and bias force of the next link from the kinematic chain. The ABA algorithm starts computing the articulated inertia of the tip body for which $\hat{\bm{I}}^A_i=\hat{\bm{I}}_i$  and then propagates it towards the parent bodies. Step 3 consists then in a tip-to-root loop. 



\subsubsection*{Step 4: Computation of the joint and link acceleration}

The time derivation of equation \eqref{eq_vi} results in the following expression for the acceleration of link $i$: 

\begin{equation}\label{eq_ai}
    \hat{\bm{a}}_i  =  {}_{i} \hat{\bm{X}}_{p(i)} \hat{\bm{a}}_{p(i)} + \hat{\bm{c}}_i +  \hat{\bm{\Phi}}_{i} \ddot{q}_i,
\end{equation}
where $\ddot{q}_i$ can be formulated from equations \eqref{eq_fpi}, \eqref{eq_pA} and \eqref{eq_Qi}:

\begin{equation} \label{eq_qi}
    \ddot{q}_i = \frac{Q_i - \hat{\bm{\Phi}}^T_i  \hat{\bm{I}}^{A}_i {}_{i} \hat{\bm{X}}_{p(i)} \hat{\bm{a}}_{p(i)} - \hat{\bm{\Phi}}^T_i (\hat{\bm{p}}^{A}_i  + \hat{\bm{I}}^{A}_i\hat{\bm{c}}_i ) }{\hat{\bm{\Phi}}^T_i  \hat{\bm{I}}^{A}_i \hat{\bm{\Phi}}_i} .
\end{equation}
These equations form the final loop of the ABA algorithm. The tree is visited again from its root to its tip to compute each joint and link acceleration.  

\subsubsection*{Step 5, 6 and 7: Numerical integration and system correction}

Following the computation of $\ddot{q}_i$ for time step $t_k$, the joint acceleration update can be relaxed based on the previous time step to enhance numerical stability. 
The resulting accelerations are integrated to compute the velocities at the next time step $t_{k+1}$ (defining $\Delta t=t_{k+1}-t_{k}$). Similarly, the joint velocities are integrated to compute the joint positions at $t_{k+1}$ and the system state $\dot{q}_i(t_{k+1})$,$q_i(t_{k+1})$ serves as the initial conditions for the equations of motion for the next iteration of the forward dynamics (step 0).

\color{black}
OpenFOAM supports three second-order integration schemes: Newmark (implicit) \cite{Newmark1959}, Symplectic (explicit) \cite{Dullweber1997}, and Crank–Nicolson (implicit) \cite{Crank1947}. Since the Crank–Nicolson scheme is a specific case of the Newmark method for particular parameter choices, this work focuses on the Newmark and Symplectic schemes.

The Newmark method updates the joint velocities and positions as:
\begin{align}\label{eq:newmark}
       \dot{q}_{i}(t_{k+1}) &= \dot{q}_{i}(t_{k}) + \Delta t \big(\gamma\ddot{q}_{i}(t_{k+1}) +  (1-\gamma)\ddot{q}_{i}(t_{k})\big) \\
          q_{i}(t_{k+1}) &= q_{i}(t_{k}) + \Delta t \dot{q}_{i}(t_{k}) + \Delta t^2 \big(\beta\ddot{q}_{i}(t_{k+1}) + (1-\beta)\ddot{q}_{i}(t_{k})\big),
\end{align}
where $\gamma,\beta$ are the hyperparameters set to 0.5 and 0.25 by default.
The implicit nature of the method requires $nIter$ updates of positions and velocities within each time step by repeating steps 0 to 7 in Figure \ref{fig:diagram}. These body state updates enhance solver stability, which is particularly important for stiff and strongly coupled problems.
Importantly, the flow equations are not solved during these $nIter$ iterations, keeping the flow forces constant (step 0). This approach reduces the computational cost but limits the algorithm's applicability to moderately coupled problems where added mass effects are weak. The drones investigated in this work have smooth-wing kinematics which generate only negligible added mass effects, especially in air. 

The second integration scheme considered is the explicit Symplectic scheme. Only one integration is performed per time step and the time steps must be kept small to maintain numerical stability in strongly coupled problems. The Symplectic scheme uses the leapfrog method for which a half step is taken to compute the joint velocities and positions:

\begin{align}\label{eq:symplectic1}
    \dot{q}_{i}(t_{k+0.5}) &= \dot{q}_{i}(t_{k}) + 0.5\Delta t\ddot{q}_{i}(t_{k-1}) \\
   q_{i}(t_{k+1}) &= q_{i}(t_{k}) + \Delta t \dot{q}_{i}(t_{k+0.5}).
\end{align}

The system dynamics is then solved (steps 0 to 4) and another half step is taken to update the joint velocities:
\begin{equation}\label{eq:symplectic2}
    \dot{q}_{i}(t_{k+1}) = \dot{q}_{i}(t_{k+0.5}) + 0.5\Delta t\ddot{q}_{i}(t_{k}),
\end{equation}

Following the updates of the joint positions and velocity using Symplectic or Newmark integration, the link position, velocity, and acceleration are corrected in step 7. The grid must then be adapted to this new state (Figure \ref{fig:diagram}).
\color{black}

\subsection{Grid adaptation and PIMPLE loop}\label{sec:gridAdaptation}
Referring to the second block in Figure \ref{fig:diagram}, the body displacement is computed from the current position at $t_{k+1}$ and the previous position at $t_{k}$. The displacement is then applied to the grid cells with the \textit{pointDisplacement} field. The cell centers of the grid are adjusted and the boundary conditions are corrected accordingly.

This work uses the overset method to adapt the grid to the body's motion. A small \textit{component} grid is fitted to each body of the system, overlapping with a large \textit{background} grid. The \textit{component} grids can also partially overlap with each other, and interpolations allow the grids to exchange velocity and pressure fields according to a role attributed to each cell. 
\textcolor{black}{
Active cells are standard cells where the flow equations are solved. 
Hole cells are background grid cells that overlap with the body and must block the flow. Interpolated cells are either neighbouring cells from the holes or cells at the component grid interface. They receive flow variables from interpolation stencils formed by active cells from another grid. Identifying these donor cells and the holes is computationally intensive as it requires spanning the background and component grid cells. Moreover, this operation must be repeated at the beginning of each time step. The reader is referred to \cite{Hadzic2006,ofWiki} for more details.}
Once the grid is updated and the fluxes are corrected according to the grid's motion, the momentum and pressure equations are solved to compute a new pressure field (block three in Figure \ref{fig:diagram}). The surface forces (equation \eqref{eq_fe}) are updated, and the full loop starts again with the rigid body dynamics solver.

\section{Implementations in the rigid body dynamics library}\label{sec:impl}

The articulated-body algorithm, \textcolor{black}{available in OpenFOAM v2206}, solves the equations of motion to predict the kinematics of the joints and links. However, various multibody systems have joints that are actively controlled by actuators. For example, an actuator controls the rudder of a boat, the elevation flaps of an airplane, and the wings of a flapping wing drone, while the rest of the system dynamics is governed by the equations of motion.

The \textit{rigidBodyDynamics} library partially allows modeling those constrained motions through a \textit{restraint} named \textit{prescribedRotation}. The latter controls the rotational velocity $\bm{\omega}$ of a given link $i$ by adding a force $\hat{\bm{f}}^p_{i}$ to its external force vector $\hat{\bm{f}}^e_{i}$. This additional force is modeled as the output of a Proportional-Integral-Derivative (PID) controller which seeks to minimize the rotational velocity error $\bm{e}_\omega(t_k) = \bm{\omega}(t_k) - \bm{\omega}_{set}(t_k)$:

\begin{equation}
    \bm{f}^p(t_k) = \frac{I_i }{dt} \Big(  K_p \bm{e}_\omega(t_k)  + K_d (\bm{e}_\omega(t_k) - \bm{e}_\omega(t_{k-1})) + K_i \sum_{t_j = {t_0}}^{t_k} \bm{e}_\omega(t_j) \Big),
\end{equation}
where $K_p, K_d, K_I$ are constant PID parameters and $dt$ is the simulation time step. 
The \textit{prescribedRotation} restraint could be easily generalized to control the motion defined by any joint (hence not only the rotation velocity of links). 
Nevertheless, the use of a PID controller requires tuning three parameters for each control scenario. This calibration process is tedious and unrealistic for computationally expensive simulations. \\


This work proposes to address these limitations by directly enforcing the kinematics $q,\dot{q},\ddot{q}$ of active joints from a multibody system, bypassing the need to compute unknown joint forces and torques.
Inspired by the work in \cite{Docquier2013,Hu2005,Featherstone2014}, the proposed implementations modify the equations of motion to make the multibody system dynamics consistent with the prescribed motion of a subset of joints. The implementations consist of three main steps in the \textit{rigidBodyDynamics} library. (1) The position, velocity, and acceleration of the active joints are loaded from user-defined analytical expressions, (2) the position, velocity, and acceleration are imposed on the desired joints, and (3) the system equations of those joints are modified to model their constrained dynamics while the equations of motion for the passive joints remain unchanged.

\begin{minipage}{0.45\linewidth}
 \begin{lstlisting}[emph={ddt,div,laplacian},caption={Snippet of \textit{wangParametrization.C}.},captionpos=b,label=code:wangC,basicstyle=\tiny]
(...)
void Foam::RBD::ImposedMotion::
wangParametrization::loadImposedMotion
(
   Field<label>& jointIndex,
   Field<scalar>& imposedJoints
) const
{
if (motionType_ == "midStart") 
{
    (...)
    Field<scalar> varList(
    jointList_.size()*3);
    
    scalar q_phi = Aphi_/asin(Kphi_)* 
    asin(Kphi_*sin(2*M_PI*fphi_*ti)); 
    scalar qDot_phi =  (...)
    scalar qDdot_phi = (...)
    
    varList[0] = q_phi; 
    varList[1] = qDot_phi;
    varList[2] = qDdot_phi; 
    
    jointIndex = jointList_;
    imposedJoints = varList;
} 
(...)

\end{lstlisting}  
\end{minipage}
\hfill
\begin{minipage}{0.45\linewidth}
 \begin{lstlisting}[emph={ddt,div,laplacian},caption={Snippet of \textit{wangParametrization.H}.},captionpos=b,label=code:wangH,basicstyle=\tiny]
class wangParametrization
: public imposedmotion
{
// Parameters read from dynamicMeshDict
    labelList jointList_;
    string motionType_;
    
    scalar Aphi_;
    scalar fphi_;
    (...)   
public:
    (...)
    
//- Compute the q,qd,qdd from joints in jointIndex and 
// save it in imposedJoints  
// [q0,qd0,qdd0,q1,qd1,qdd1,...]
    virtual void loadImposedMotion
    (
        Field<label>& jointIndex,
        Field<scalar>& imposedJoints
    ) const;
    
//- Update properties from given dict
    virtual bool read(const
    dictionary& dict);
    
//- Write
    virtual void write(Ostream&) const;
(...)

\end{lstlisting}   
\end{minipage}

\subsubsection*{Implementation 1}
A new library named \textit{imposedMotion} is implemented based on the existing \textit{restraint} library.
The \textit{imposedMotion} class defines generic operations that child classes can leverage, with each child class specifying different motion parameterizations, as demonstrated in Listings \ref{code:wangC} 
 and \ref{code:wangH} for \textit{wangParametrization}.

The class defines private variables read from the \textit{dynamicMeshDict} dictionary. \texttt{joinList$\_$} contains the indices of the active joints and \texttt{motionType$\_$} designates the parameterization subtype implemented within the \texttt{loadImposedMotion} member function.
This function saves the positions, velocities, and accelerations in \texttt{imposedJoints} from user-defined analytical expressions for the joints with indices in \texttt{joinList$\_$}. The analytical expression is a function of the current time and parameters defined in the header.


\subsubsection*{Implementation 2}
After the integration of the acceleration and the velocity (e.g. in \textit{Newmark.C}), a loop iterates on the \texttt{jointIndex} list and imposes the $q$, $\dot{q}$ and $\ddot{q}$ from \texttt{imposedJoints} to the corresponding joints. The same implementation is done for the three numerical integration schemes supported by OpenFOAM (i.e. \textit{Newmark.C}, \textit{symplectic.C}  and \textit{CrankNicolson.C}).\\

\subsubsection*{Implementation 3}
The last implementation modifies the equations of motion to make the dynamics of the multibody system consistent with the imposed motions. 
Changes take place in steps 3 and 4 of the ABA algorithm implemented in $forwardDynamics.C$ (Figure \ref{fig:diagram}).
For active joints, the acceleration $\ddot{q}$ in equation \eqref{eq_qi} is known with the implementations 1 and 2. 
The equation can be manipulated to isolate the magnitude of the unknown joint force $Q_i$ that drives the motion of the joint. The latter is inserted in equation \eqref{eq_pA}, which changes the expressions of the  articulated inertia and the bias force of link $p(i)$ as:

\begin{equation}\label{eq_IAimposed}
\hat{\bm{I}}^{A}_{p(i)} = \hat{\bm{I}}_{p(i)} +   {}_{p(i)}\hat{\bm{X}}_{i} \hat{\bm{I}}^{A}_i {}_{i}\hat{\bm{X}}_{p(i)},
\end{equation}

\begin{equation}\label{eq_pAimposed}
\hat{\bm{p}}^A_{p(i)} = \hat{\bm{p}}_{p(i)} +   {}_{p(i)}\hat{\bm{X}}_{i} \Big( \hat{\bm{p}}^{A}_i +  \hat{\bm{I}}^{A}_i \hat{\bm{c}}_i + \hat{\bm{I}}^{A}_i \hat{\bm{\Phi}}_i \ddot{q} \Big).
\end{equation}

\textcolor{black}{
The resulting extended articulated-body algorithm (eABA) is summarised with the Algorithm \ref{alg:ABA} for which the red colour highlights the new implementations compared to the original ABA (in black). 
Line 1 calculates external loads using Equation \eqref{eq_fe} and user-defined forces/torques specific to the multibody test case. User-defined joint forces $Q_j$ are also calculated, except for the $n_a$ active joints driven by analytical functions in the eABA implementation.
Line 2 loads the $n_a$ analytical expressions of the position, velocity and acceleration (solely in eABA).
Lines 5–8 (common to both ABA and eABA) calculate link quantities needed for inertia and bias force computation (lines 9–14). A conditional statement is added in the eABA to use equations \eqref{eq_IAimposed} and \eqref{eq_pAimposed} for active joints (Line 12) and use equations \eqref{eq_IA} and \eqref{eq_pA} otherwise. Lines 15–18 compute joint and link accelerations. In the eABA algorithm, acceleration computation of the active joints (equation \eqref{eq_qi}) is not performed (line 16)  since the accelerations, velocities and positions are imposed later (line 22) using the expression loads in line 2. With these implementations, the free motion of multibody systems can be simulated in a flow while one of its bodies follows a known trajectory. This is verified for the flapping-wing drones defined in the next section.
}


\begin{algorithm}
\caption{Extended articulated-body algorithm using Newmark time integration and with changes compared to the ABA indicated in red}\label{alg:ABA}
Compute the external loads and restraints in $\hat{\bm{f}}^e_j$ and $Q_j$  for the $n\textcolor{red}{-n_a}$ links \\
\textcolor{red}{Load the indices of the active joints in \textit{jointIndex} and the imposed motion in $\bm{q}^a,\dot{\bm{q}}^a,\ddot{\bm{q}}^a$}\\
Set the root velocity to 0 and its acceleration to the gravity \\
\CommentSty{\textbf{Forward dynamics}}\\
\Indp \For{\text{link $i$ = root to tip}}{
Compute link velocities $\hat{\bm{v}}_i$: equation \eqref{eq_vi} \\
Compute velocity-product forces $\hat{\bm{c}}_{i}$: equation \eqref{eq_c}\\
Compute isolated inertia $\hat{\bm{I}}_{i}$ and bias force $\hat{\bm{p}}_{i}$: equation \eqref{eq_Ii} and \eqref{eq_pI}\\
}

\For{\text{link $i$ = tip to root}}{
Compute articulated-body inertia $\hat{\bm{I}}^{A}_{i}$ and bias forces $\hat{\bm{p}}^A_{i}$ with: \\
  \textcolor{red}{\uIf{link's joint in jointIndex}{
    Equation \eqref{eq_IAimposed} and \eqref{eq_pAimposed} 
  }}
  \uElse{
    Equation \eqref{eq_IA} and \eqref{eq_pA}
    }
}
\For{\text{link $i$ = root to tip}}{
  \uIf{\textcolor{red}{link's joint not in jointIndex}}{
   \textcolor{black}{Compute joint acceleration: Equation \eqref{eq_qi}}
  }
  Compute link acceleration: Equation \eqref{eq_ai}
}
\Indm\CommentSty{\textbf{Time integration}} \\
\Indp\For{\text{joint $i$ = root to tip}}{
  \textcolor{red}{\uIf{joint $i$ in jointIndex}{
   Impose joint kinematics $q_i = q_i^a$, $\dot{q}_i = \dot{q}_i^a$, $\ddot{q}_i = \ddot{q}_i^a$ 
  }}
    \uElse{
    $q_i,\dot{q}_i,=\text{Newmark}(\dot{q}_i,\ddot{q}_i)$ 
    }
}
\end{algorithm}

\section{Test case definition}
Two test cases are defined in the following section. The first test case is a single-wing (SW) drone used to verify the rigid body dynamics CFD environment in section \ref{sec:result_1}. The second test case is a body and wing (BW) drone used to analyze the dynamics of an ascending flight in section \ref{sec:result_2}. In both test cases, only one wing is considered, defining the degrees of freedom to ensure symmetry of the flapping motion to the ($z_I,x_I$) plane.

\subsection{Single-wing (SW) drone} \label{sec:case1}
\begin{figure}[!h]
\includegraphics[width=\textwidth]{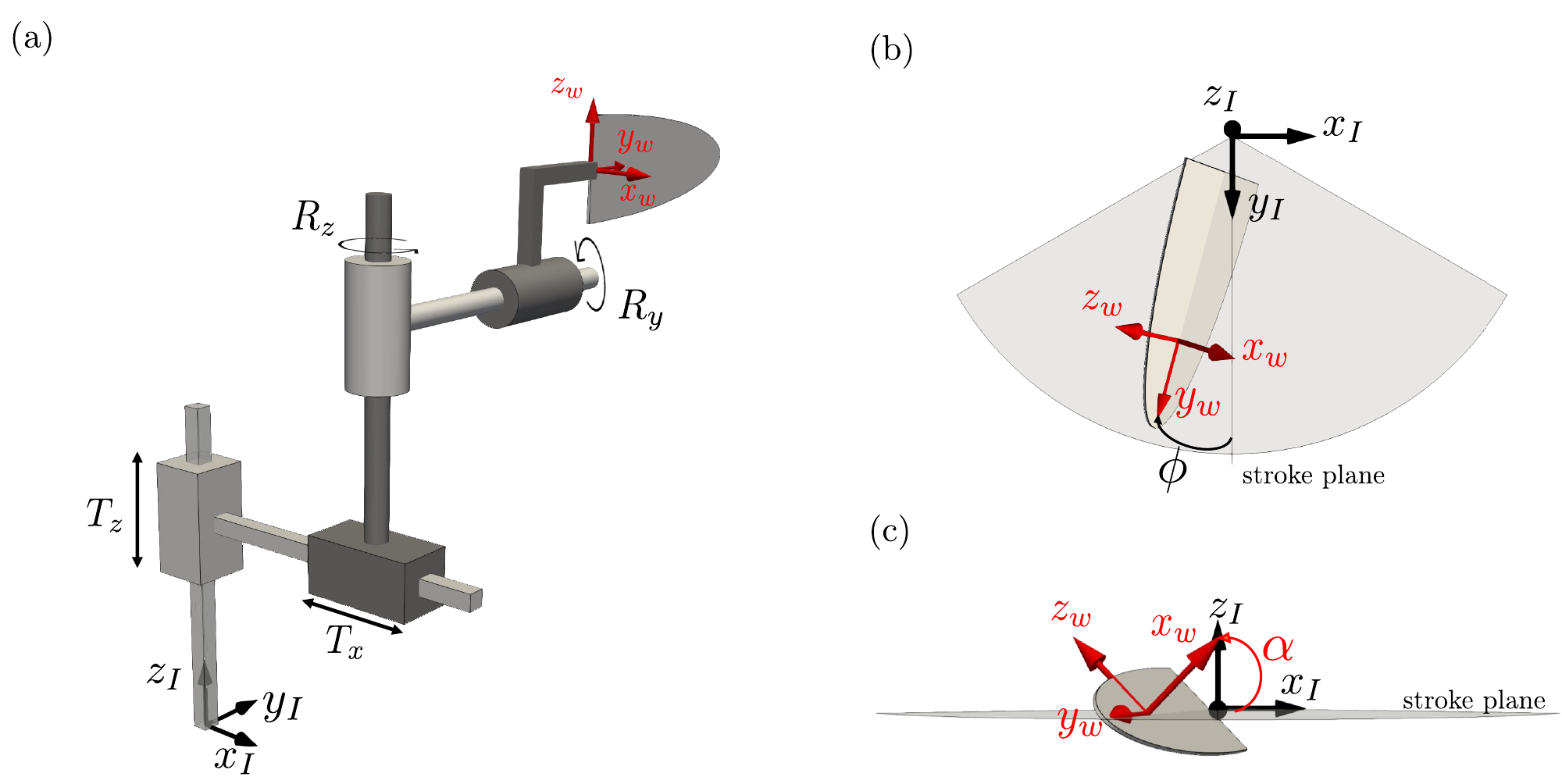}
\caption{\textcolor{black}{(a) Kinematic tree of the flapping-wing drone, showing its four degrees of freedom. (b) Top view and (c) side view of the wing, defining the flapping angle ($\phi$) and pitching angle ($\alpha$).}}
\label{fig:wing_full}
\end{figure}

The SW drone consists of a massless body and rigid, semi-elliptical wings, as shown in Figure \ref{fig:wing_full}. 
\textcolor{black}{The wing weighs half of a typical hummingbird \cite{Cheng2016,altshuler2012,Song2014}} with a centre of mass that coincides with the instantaneous centre of rotation. The centre of mass is then outside the wing geometry, in a position similar to that of a hummingbird. \textcolor{black}{This SW drone assumes then a negligible inertial and aerodynamic body-wing coupling as discussed in the introduction and in \cite{Taha2012,Orlowski2011}. This hypothesis greatly simplifies the verification analyses undertaken in section \ref{sec:result_1} but may result in unrealistic flight trajectories as verified with the BW drone test case (section  \ref{sec:case2}).}


\textcolor{black}{The wing geometry is defined according to our previous works \cite{poletti2024} using the design from \cite{Lee2016}, inspired by large hovering flyers such as hummingbirds \cite{Chin2016}.}
Table \ref{tab:parameters} gathers its main dimensions, and Figure \ref{fig:wing_full}.a shows its kinematic chain. The chain begins with two translational joints ($T_{z_I},T_{x_I}$) along the vertical and longitudinal directions defined in the inertial frame $(x,y,z)_I$. \textcolor{black}{These are used to solve for the drone trajectory defined with the variables $x$ and $z$.}
The prismatic joints are followed by two revolute joints ($R_{z_I},R_{y_w}$) around $z_I$ of the inertial frame and $y_w$ of the wing frame $(x,y,z)_w$ that coincides with the wing's axis of symmetry. These are used to prescribe the wing kinematics. Thus, the full kinematic chain conveys four degrees of freedom to the wing: the translational positions are computed by solving the equations of motion, while the rotational angles are imposed to follow periodical motions. \textcolor{black}{The popular "Wang" parametrization \cite{Berman2007,Yan2015,Bayiz2018,Bhat2020} is chosen since it allows to impose periodical waveforms to the revolute joints using only two parameters.} It is formulated as:

\begin{align}
\label{eq:phi}
\phi (t) &= \frac{A_{\phi}}{\arcsin (K_{\phi})}\arcsin [K_{\phi}\cos(2 \pi  ft)], \\
\label{eq:alpha}
\alpha (t) &= \frac{A_{\alpha}}{\tanh (K_{\alpha})}\tanh [K_{\alpha}\sin(2 \pi  ft)], \
\end{align}
where equation \eqref{eq:phi} defines the flapping angle due to the $R_{z}$ joint and equation \eqref{eq:alpha} defines the pitching angle due to the $R_{y}$ joint.
Figure \ref{fig:wing_full}  shows both angles in the stroke plane parallel to $(x_I,y_I)$  and in which the tip of the wing always remains.
$A_\phi, A_\alpha $ are the maximum flapping and pitching angles, $f$ is the frequency for both rotations and $K_\phi$ and $K_\alpha$ drive the flapping and pitching waveforms. This article considers sinusoidal waveforms with the $A_\phi$, $A_\alpha$ and $f$ from table \ref{tab:parameters} that were inspired by experimental measurements \cite{Ellington1984_III,Kruyt2014} and CFD simulations \cite{Bos2013,Nakata2015,Song2014} on the hovering flight of natural species.

In OpenFOAM, the degrees of freedom are defined in the \textit{dynamicMeshDict} wherein the \textit{imposedMotion} dictionary with the \textit{wangParametrization} is called to enforce the rotational motions (see Section \ref{sec:impl}). The Symplectic scheme is used to integrate the equations of motion and compared with the Newmark solver in section \ref{sec:parametricStudies}. The integration is done once three times per time step without acceleration relaxation (see Figure \ref{fig:diagram} and Section \ref{sec:parametricStudies}). 

      

\begin{table}
        \centering
                \begin{tabular}{p{0.3\textwidth}p{0.1\textwidth}p{0.12\textwidth}}
                  \toprule
                          \multicolumn{3}{l}{Wing geometry}\\
                  \midrule
                          Span & $R$     & 5 cm\\
                          Offset & $R_0$     & 2.25 cm \\
                          Mean chord & $\bar{c}$     & 1.5 cm\\
                          Thickness & $b$     & 0.045 cm\\
                        \hline
                          \multicolumn{3}{l}{Wing motion}\\
                        \hline
                          Maximum flapping angle &  $A_\phi$     & 75$^\circ$ \\
                          Maximum pitching angle & $A_\alpha$      &  45$^\circ$\\
                          Frequency & $f$      &  20 Hz\\
                          Flapping shape parameter & $K_\phi$     & 0.01\\
                         Pitching shape parameter &  $K_\alpha$      &  0.01\\
                                              \hline
                          \multicolumn{3}{l}{Wing dynamic parameter}\\
                        \hline
                          Mass &  $m$     & 1.5 g \\
                          Inertia & $I_x$,$I_y$,$I_z$      &  $6e-6$ $kg\cdot m^2$\\
                          Center of mass & $\bm{x}$      &  (0,0,0) m\\
                  \bottomrule
                \end{tabular}
                \caption{Parameters of the single-wing drone test case.}
                \label{tab:parameters}
\end{table}

\textcolor{black}{
The grid is adapted to the wing motion using the overset method similar to other works focusing on tethered flights \cite{Aono2008,Badrya2016,Bomphrey2017} and free flights \cite{Liu2010,Cai2021,Bakhshaei2021}.} Two overlapping grids are defined (see Figure 2 from previous work \cite{Calado2023}). Following the guidelines in \cite{Badrya2016}, the background grid is a \textcolor{black}{20-chord} cube with 500k (predominantly) hexahedral cells that are refined along the wing path to minimize inter-grid interpolation errors \cite{Alletto2022} and ensure a detailed capture of the coherent structures generated near the wing. The interpolation scheme used is the inverse distance. 

The component grid is a C grid with 130k cells positioned within two chord lengths around the wing \textcolor{black}{following literature standards \cite{Badrya2016,Nakata2015} and previous works \cite{poletti2024}}. 
This grid follows the wing's translation and rotation, and the background grid follows the wing's translation.
This moving background grid contrasts with most of the CFD works \cite{Alletto2022,Coe2023,Ahmed2024}, which have used a fixed background grid and did not solve for the drone motion.

In the \textit{dynamicMeshDict}, the cell zone containing the background grid cells is assigned to move with the centre of gravity of the second link from the kinematic tree by using the \textit{drivenLinearMotion} library. For recall, there is a massless body between each joint, and so the second link is located just after $T_x$ (Figure \ref{fig:wing_full}).

The flow solver is \textit{overPimpleDyMFoam} which is incompressible and transient. The solver uses the PISO algorithm with four pressure correction iterations for each momentum predictor step. The flow is on the limit of the transition regime ($Re\sim\mathcal{O}(10^3)$ when the drone hovers), and the LES model with the dynamic turbulent kinetic sub-grid model was shown to give similar aerodynamic forces than a \textit{Fluent} simulation in \cite{Calado2023}

Second-order central Gauss schemes are used for the spatial terms with limiters in the direction of the most severe gradients. A backward, second-order, implicit scheme is used for the time marching, and the simulations were performed with adaptative time steps capped by a maximum Courant number of 1. 

Regarding the boundary conditions, the zero gradient condition is imposed on the boundaries of the background domain for the velocity field, together with a $0$ $m^2/s^2$ fixed value for the pressure field. The wing patch is treated as a moving wall to impose a velocity distribution that results from its motion.
The zero gradient condition is used on each domain patch for the \textit{pointDisplacement} field and the \textit{dynamictMeshDict} defines the parameters that govern the drone and the grid displacement.


\subsection{Body and wing drone}\label{sec:case2}
\begin{figure}[!h]
\centering
\includegraphics[width=0.9\textwidth]{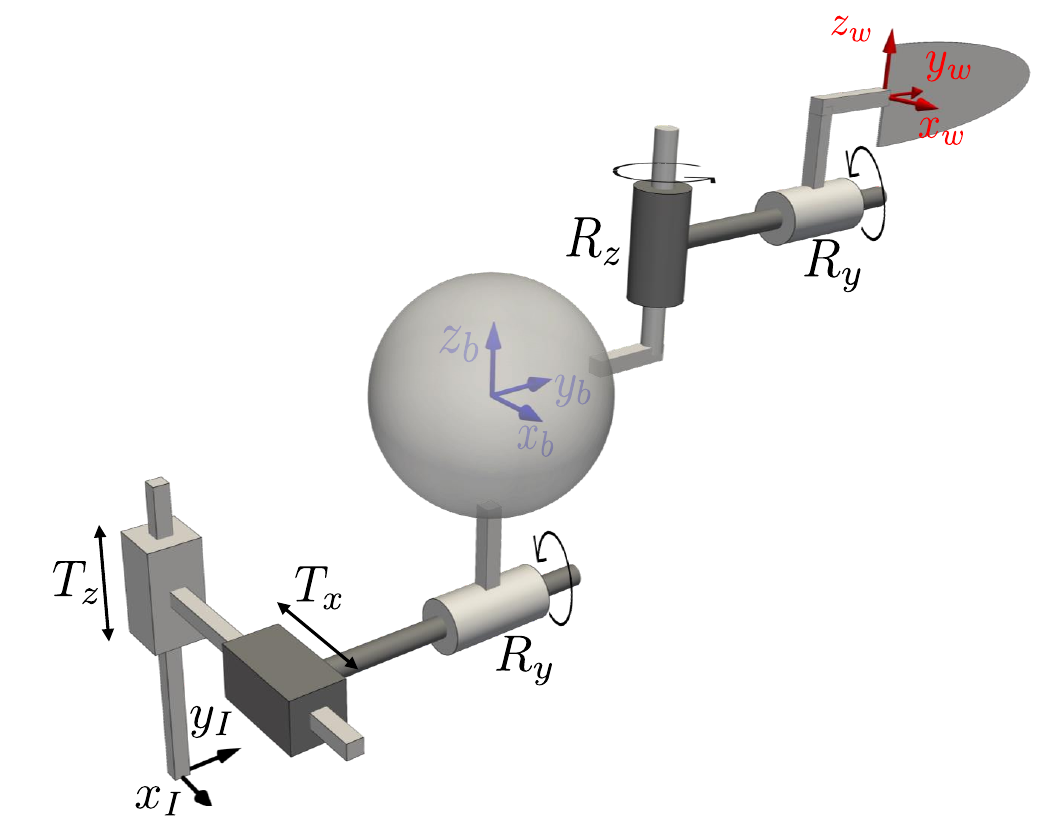}
\caption{\textcolor{black}{Kinematic tree of the drone consisting of a spherical body and a wing with 5 degrees of freedom.}}
\label{fig:MBS2}
\end{figure}

\textcolor{black}{
The second test case adds a spherical body to the single-wing drone, introducing aerodynamic and inertial coupling between the wing and the body.   
Figure \ref{fig:MBS2} shows the kinematic tree. Two translational joints ($T_z,T_x$) and one revolute joint ($R_y$) connect the spherical body to the inertial frame. The translational joints allow the horizontal $x$ and vertical displacement $z$ of the drone, and $R_y$ allows the body to pitch with angle $\theta$ as defined between $z_s$ of the body frame and $z_I$ of the inertial frame.  
The spherical body is connected to the wing by the same revolute joints used for the single-wing drone (Section \ref{sec:case1}).
The wing can execute both flapping and pitching motions relative to the body. Unlike the single-wing drone, the stroke plane is defined in $(x,y,z)_b$, which can then be tilted according to the pitching angle $\theta$.
}

\begin{figure*}[!ht]\center
\begin{subfigure}{0.45\textwidth}
    \begin{minipage}{0.05\textwidth} 
        \vspace{-7cm}
        \captionsetup{justification=raggedright,singlelinecheck=false,format=hang}
        \caption{}
        \label{fig:grid1}
    \end{minipage}%
    \begin{minipage}{0.95\textwidth} 
        \centering
        \includegraphics[width=0.99\textwidth]{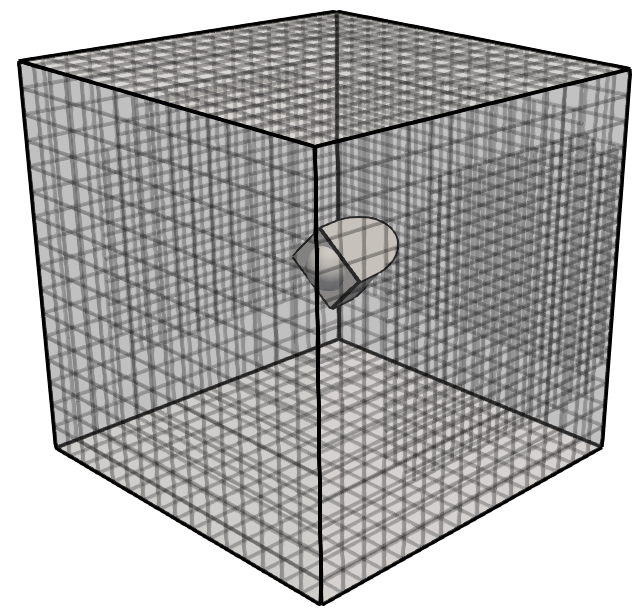}
    \end{minipage}
\end{subfigure}
\begin{subfigure}{0.45\textwidth}
    \begin{minipage}{0.05\textwidth} 
        \vspace{-7cm}
        \captionsetup{justification=raggedright,singlelinecheck=false,format=hang}
        \caption{}
        \label{fig:grid2}
    \end{minipage}%
    \begin{minipage}{0.95\textwidth} 
        \centering
        \includegraphics[width=\textwidth]{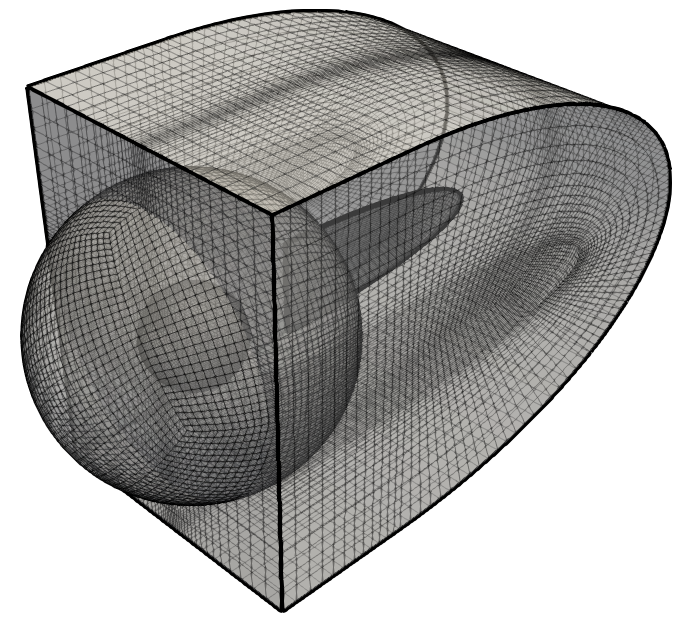}
    \end{minipage}
\end{subfigure}
\caption{(a) Background grid shown in transparent mode with the component grid of the spherical body and the wing. (b) A closer view of the component grids.}
\label{fig:oversetGrid}
\end{figure*}

The wing motion is parameterized by the equations \eqref{eq:phi} and \eqref{eq:alpha} (Table \ref{tab:parameters}). 
The sphere has a radius equal to the chord length, the wing mass is 10\% of the sphere mass and the wing inertia is 1\% of the sphere inertia (see Table \ref{tab:parameters}).
A component grid is generated around the sphere (160k cells) and the wing (130k cells) as shown in Figures \ref{fig:grid1} and \ref{fig:grid2}.   
All the other numerical parameters are set as for the single-wing drone test case. 

\textcolor{black}{
Finally, for the verifications of the CFD environment undertaken in section \ref{sec:simplifiedModels}, the dimensionless aerodynamic forces of the wing are defined in the $x_I$ and $z_I$ direction of the inertial frame as:\\ 
\begin{minipage}{0.45\textwidth}
\begin{equation}\label{eq_cxi_w}
C_{x_I,w} = \frac{\int_{S_w}(\rho p \bm{1} + \bm{\tau}_\nu) d\bm{S}_w \cdot \bm{x}_I}{0.5\rho S_wU_{ref}^2},
\end{equation}
\end{minipage}
\begin{minipage}{0.45\textwidth}
\begin{equation}\label{eq_czi_w}
C_{z_I,w} = \frac{\int_{S_w} (\rho p \bm{1} + \bm{\tau}_\nu) d\bm{S}_w\cdot \bm{z}_I }{0.5\rho S_wU_{ref}^2},
\end{equation}
\end{minipage}}

\textcolor{black}{
and similarly for the drone's body:}

\textcolor{black}{
\begin{minipage}{0.45\textwidth}
    \begin{equation}\label{eq_cxi_b}
C_{x_I,b} = \frac{\int_{S_b}(\rho p \bm{1} + \bm{\tau}_\nu) \cdot d\bm{S}_b \cdot \bm{x}_I}{0.5\rho S_bU_{ref}^2},
\end{equation}
\end{minipage}
\begin{minipage}{0.45\textwidth}
\begin{equation}\label{eq_czi_b}
C_{z_I,b} = \frac{\int_{S_b} (\rho p \bm{1} + \bm{\tau}_\nu) \cdot d\bm{S}_b\cdot \bm{z}_I }{0.5\rho S_bU_{ref}^2},
\end{equation}
\end{minipage}}

\textcolor{black}{
where the reference velocity is defined as $U_{ref}=4fA_{\phi}R_2$ at the radius of second moment of area $R_2=\sqrt{\int_{R_0}^{R_0 + R}c(r)r^2dr/S_w}$.}


\section{Results from the single-wing drone}\label{sec:result_1}
This first result section verifies the accuracy and performance of the rigid body dynamics CFD environment described in section \ref{sec:impl} using the single-wing drone test case. Section \ref{sec:parametricStudies} presents two parametric studies: a grid independence study and a study on the integration scheme. 
Section \ref{sec:simplifiedModels} compares the drone forces and trajectories with those computed by a simplified environment developed in Python. Finally, section \ref{sec:profiler} discusses the computational cost of typical simulations.

\subsection{Parametric studies} \label{sec:parametricStudies}

\begin{figure*}[!ht]\center
\begin{subfigure}{0.45\textwidth}
    \begin{minipage}{0.05\textwidth} 
        \vspace{-6cm}
      
        \captionsetup{justification=raggedright,singlelinecheck=false,format=hang}
        \caption{}
        \label{fig:xyzGrid}
    \end{minipage}%
    \begin{minipage}{0.95\textwidth} 
        \centering
        \includegraphics[width=0.99\textwidth]{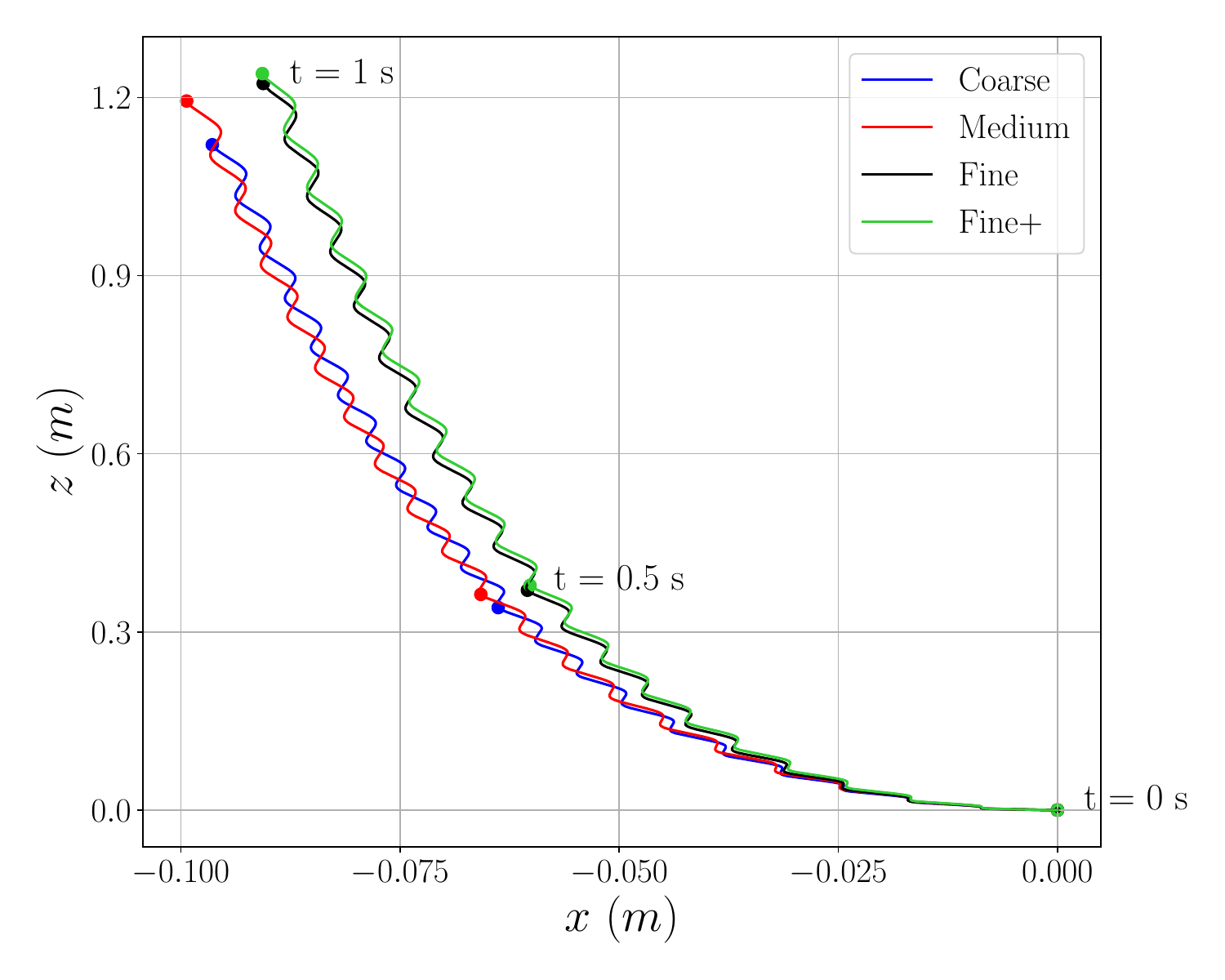}
    \end{minipage}
\end{subfigure}
\begin{subfigure}{0.45\textwidth}
    \begin{minipage}{0.05\textwidth} 
        \vspace{-6cm}
        \captionsetup{justification=raggedright,singlelinecheck=false,format=hang}
        \caption{}
        \label{fig:xyzCFL}
    \end{minipage}%
    \begin{minipage}{0.95\textwidth} 
        \centering
        \includegraphics[width=\textwidth]{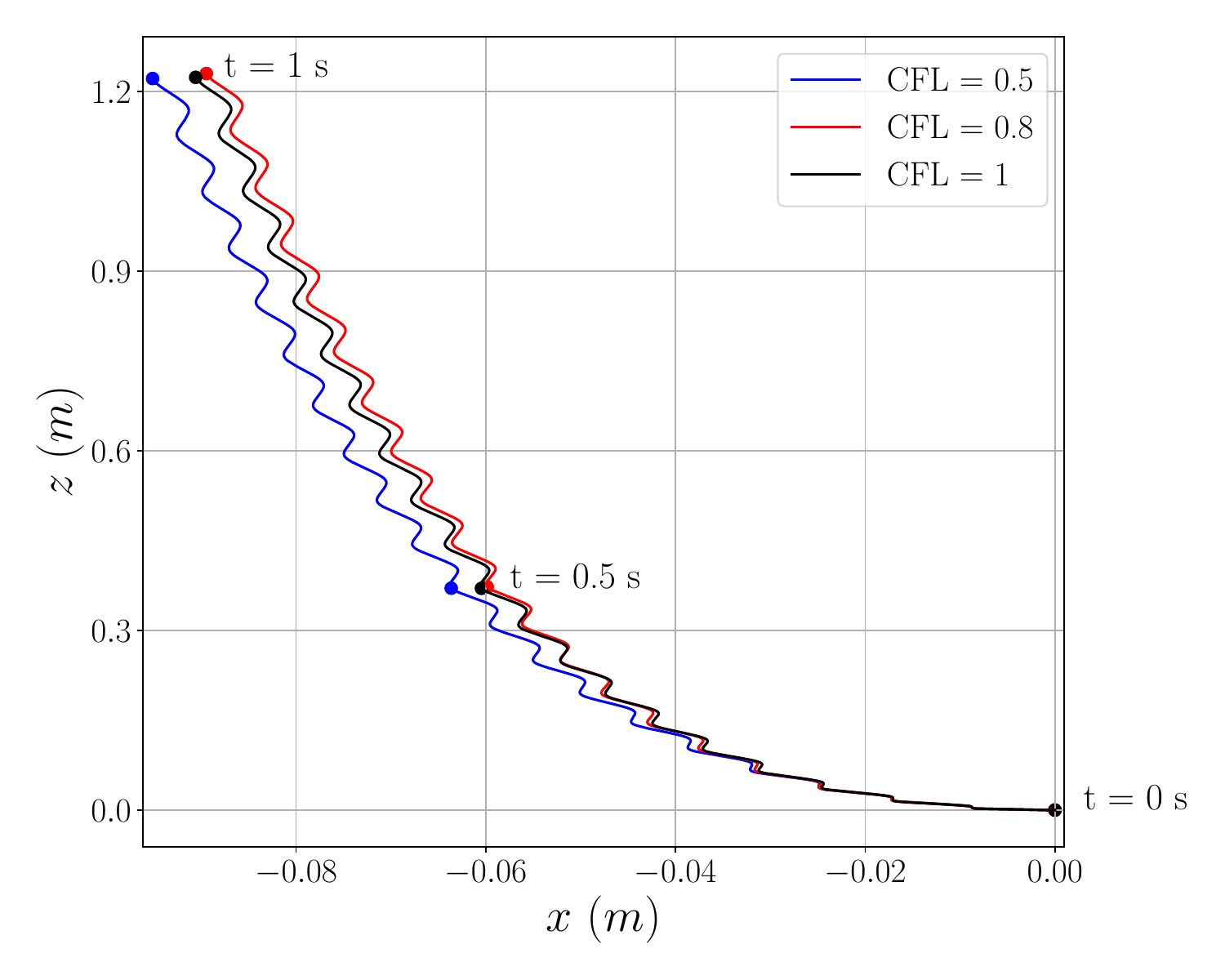}
    \end{minipage}
\end{subfigure}
\begin{subfigure}{0.45\textwidth}
    \begin{minipage}{0.05\textwidth} 
        \vspace{-5.5cm}
        \captionsetup{justification=raggedright,singlelinecheck=false,format=hang}
        \caption{}
        \label{fig:xyzIntegration}
    \end{minipage}%
    \begin{minipage}{0.95\textwidth} 
        \vspace{0.5cm}
        \centering
        \includegraphics[width=\textwidth]{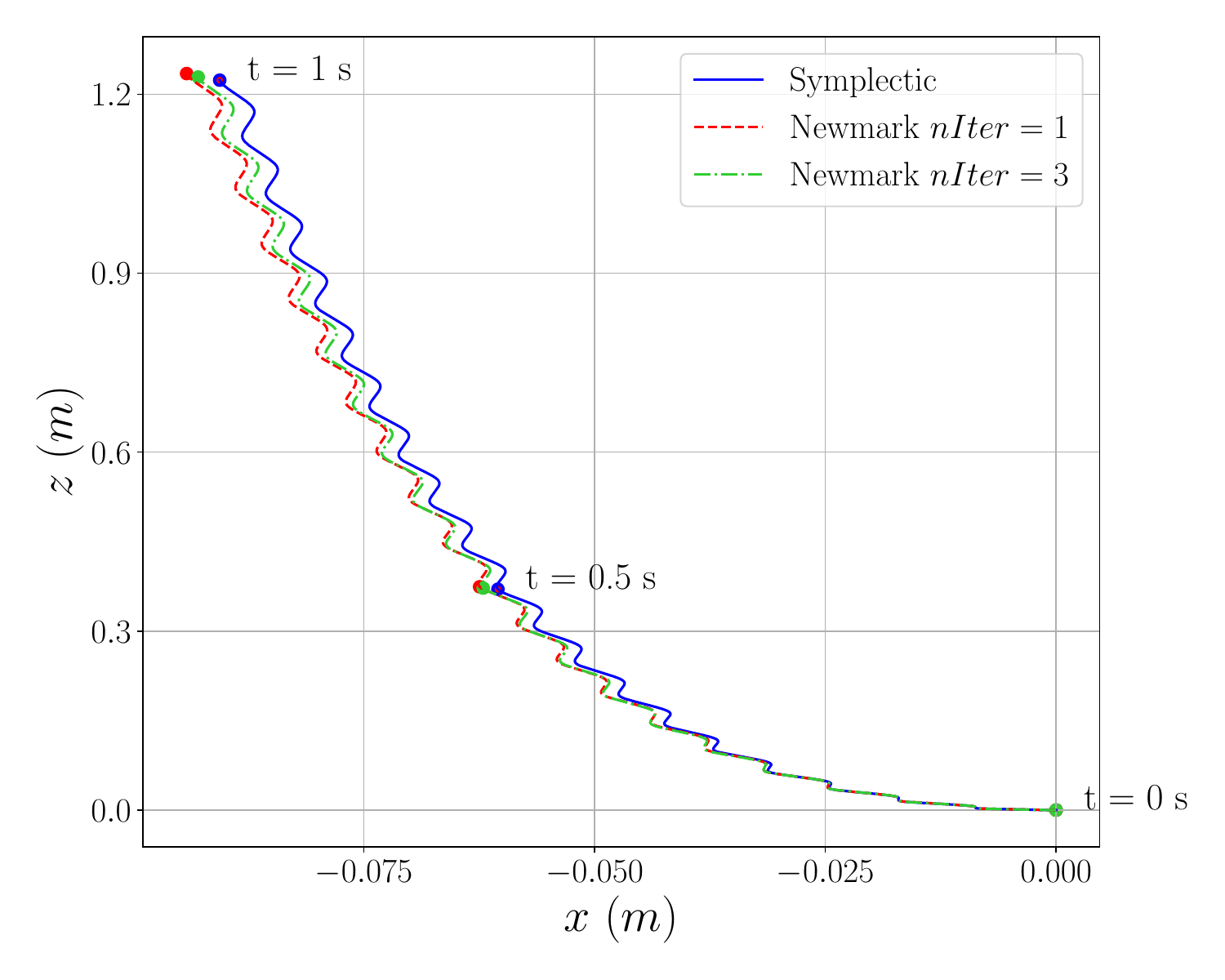}
    \end{minipage}
\end{subfigure}
\caption{(a) x-z trajectories for a coarse (71k cells) medium (214k cells), fine (501k cells), and fine+ (808k cells) grid, \textcolor{black}{(b) for CFL = 0.5, 0.8 and 1} and (b) for the Newmark and Symplectic integration schemes.}
\label{fig:numerics_full}
\end{figure*}

Figure \ref{fig:xyzGrid} tests the influence of the grid refinement on the x-z trajectory of the drone. 
The grids are refined by first increasing the cell count of the component grid and then adapting the grid refinement of the background grid to conserve similar cell sizes at the interface between the two grids. The influence of the grid on the trajectory is small and assessed with the error $e = (||\bm{x}_f(i)||_2-||\bm{x}_f(i+1)||_2)/||\bm{x}_f(i+1)||_2$ for which $i\in\{Coarse, Medium, Fine, Fine+\}$ and $\bm{x}_f=(x,z)$ is the final position taken when $t=1$ s.
A 6.1\%, 2.3\% and 1.3 \% is computed going from the coarsest to the finest grid. One can assume that sufficient grid independence is reached with the fine grid. 
\color{black}
Similarly, Figure \ref{fig:xyzCFL} shows the influence of the step size $\Delta t$ imposed through the CFL number. The average $\Delta t$ is $7.7e-06$ $s$ for CFL = 0.5 and the double CFL = 1. The figure clearly shows that the step size with CFL = 1 is sufficiently small to result in trajectories that are almost overlapping with those with CFL = 0.5.
\color{black}

\color{black}
Figure \ref{fig:xyzIntegration} shows the influence of the numerical scheme used to integrate the equations of motion. The symplectic method (equation \eqref{eq:symplectic1} and \eqref{eq:symplectic2}) is compared with the Newmark method (equation \eqref{eq:newmark}), evaluated with one and three passes of the rigid body dynamics solver per time step. 
Both methods rely on similar time steps determined by the adaptive time-stepping scheme of the flow solver. 

The figure demonstrates that both solvers produce similar trajectories, with the final position differing by less than $1\%$. It also shows that increasing sub-iterations with the Newmark scheme brings the trajectory closer to that of the symplectic method. Notably, no stability issues were observed with either method.
\color{black}

\subsection{Verification studies}  \label{sec:simplifiedModels}
\color{black}
Parametric studies from the previous section demonstrated the robustness of the CFD setup, showing that the results are not sensitive to numerical parameters. The next step in the validation process is to benchmark the CFD results, ideally with experimental data. 
However, flapping-wing drones are rare, and none offer sufficiently reproducible setups or detailed results. Similarly, numerical simulations of flapping-wing flights remain scarce, as noted in the introduction. Few studies have employed multibody formulations but they rely on complex setups, making them impractical for validation purposes.
Given these limitations, the CFD environment is verified in three steps.

\subsubsection{Flow solver validation} 
The flow solver was first validated without the body dynamics solver in a previous work \cite{poletti2024} (using the overset method and LES).  Harmonic motions of the wing were simulated in quiescent flow and the aerodynamic forces were found to match the ones from a Fluent simulation using the sliding mesh approach \cite{Lee2016}. 

\subsubsection{Rigid body dynamic solver validation}
The rigid body dynamic solver (eABA) is validated independently from the flow solver using a 2D double pendulum in a vacuum. This multibody system is a standard benchmark, extensively analyzed in previous numerical studies \cite{Stachowiak2006,Calvao2015,Herho2024}.
Its non-linear and chaotic dynamical system is defined using Lagrangian mechanics and implemented in Python following \cite{Herho2024}.
The resulting equations of motion are time-integrated using a Runge-Kutta scheme to provide the benchmark data for comparison against the double pendulum simulated within the OpenFOAM environment.

\begin{figure*}[!ht]\center
\begin{subfigure}{0.45\textwidth}
    \begin{minipage}{0.05\textwidth} 
        \vspace{-6cm}
      
        \captionsetup{justification=raggedright,singlelinecheck=false,format=hang}
        \caption{}
        \label{fig:pend}
    \end{minipage}%
    \begin{minipage}{0.95\textwidth} 
        \centering
        \vspace{0.5cm}
        \includegraphics[width=0.99\textwidth]{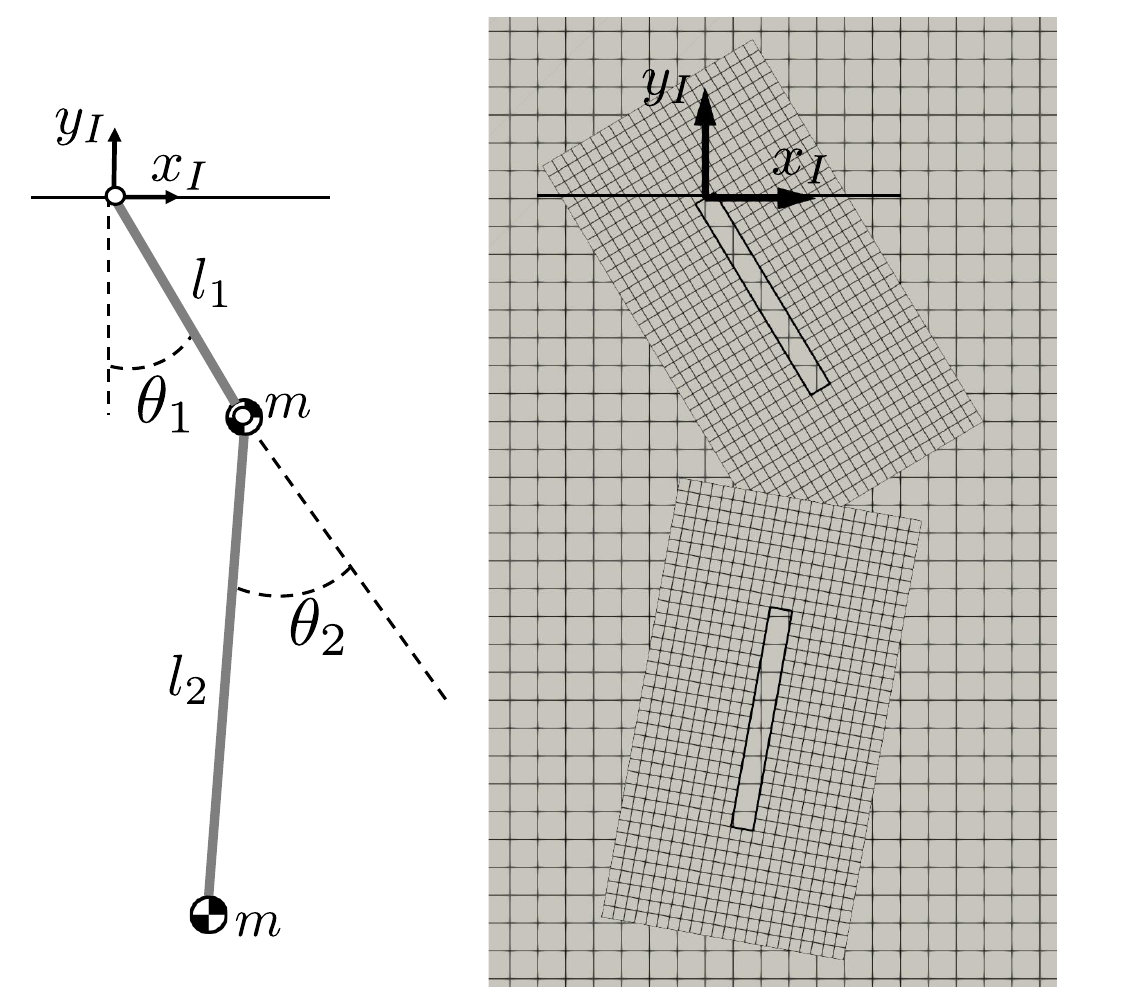}
    \end{minipage}
\end{subfigure}
\begin{subfigure}{0.45\textwidth}
    \begin{minipage}{0.05\textwidth} 
        \vspace{-6cm}
        \captionsetup{justification=raggedright,singlelinecheck=false,format=hang}
        \caption{}
        \label{fig:pend_res}
    \end{minipage}%
    \begin{minipage}{0.95\textwidth} 
        \centering
        \vspace{0.5cm}
        \includegraphics[width=\textwidth]{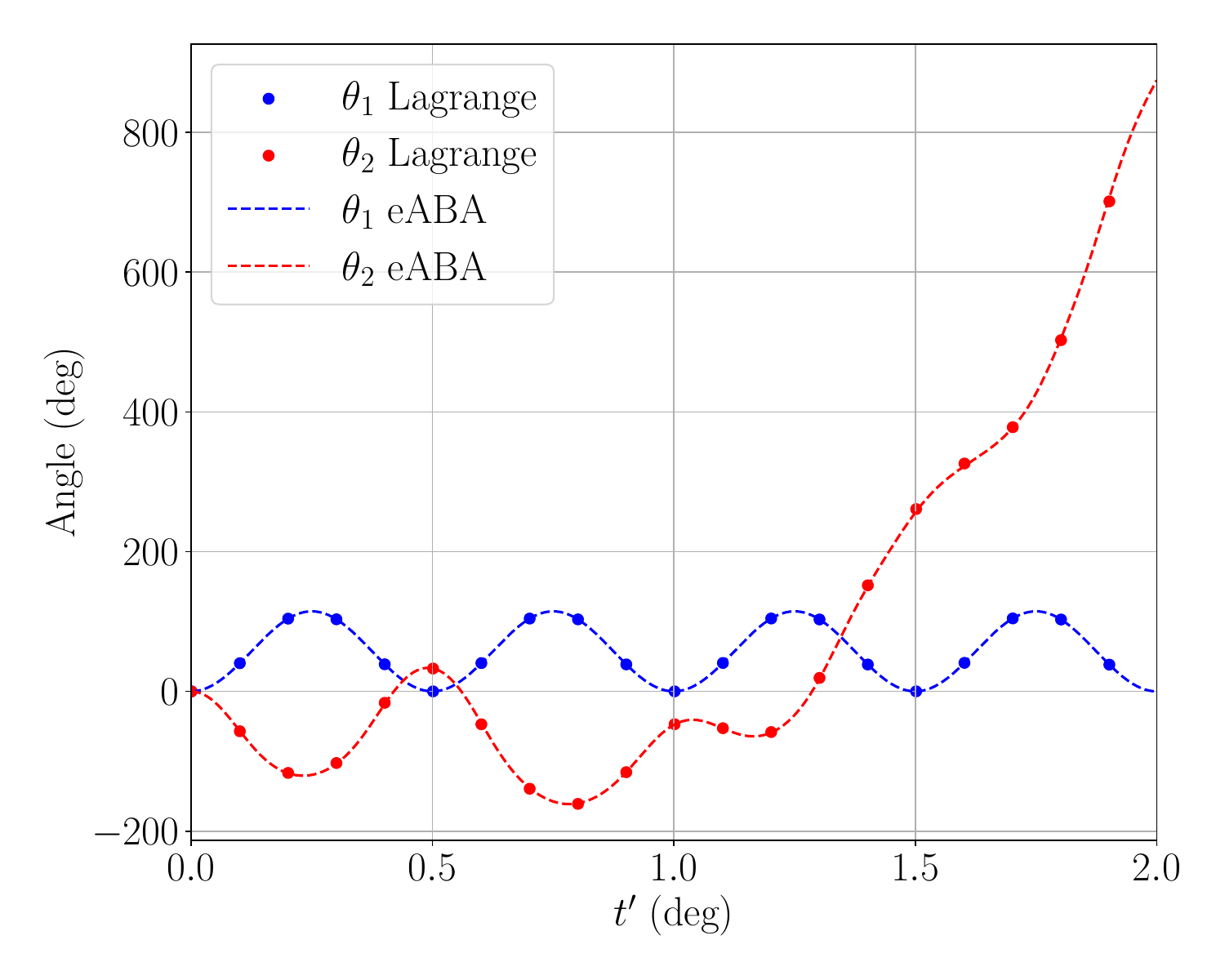}
    \end{minipage}
\end{subfigure}
\caption{(a) Schematic of the double pendulum, its two component grids and the background grid and (b) Comparison of the pendulum angles $\theta_1$ and $\theta_2$ using the eABA and Lagrange mechanics. }
\label{fig:pendAll}
\end{figure*}

The double pendulum is shown in Figure \ref{fig:pend}. It consists of two massless rods with length $l_1=10$ cm and $l_2=20$ cm, each with a point mass $m=0.5$ $kg$ at its end.
A revolute joint $R_z$ connects the top of the first rod to the origin of the inertial frame $(x,y)_I$. A second revolute joint $R_z$ connects the tip of the first rod with the top of the second rod. The system motion is then characterized by the angles $\theta_1$ and $\theta_2$. Figure \ref{fig:pend} (right) illustrates the grid setup, where the component grid for the second rod is generated only around its second half. This design ensures a clearance of $l_1$ between the two component grids, preserving a sufficient number of cells between the pendulums and preventing interpolation issues \cite{Schubert2017}.
Since no fluid forces act on the system, this computation grid setup does not influence the pendulum dynamics.

To test the eABA, the first joint of the pendulum is activated using the cosine function $f(t) = A \cos(2\pi ft) - A$. 
At each time step, this function drives the angle $\theta_1(t)=f(t)$, velocity $\dot{\theta}_1(t)=\dot{f}(t)$, and acceleration $\ddot{\theta}_1(t)=\ddot{f}(t)$, with $A=57$ (deg) and $f=2$ Hz. The resulting motion of the second rod is measured through $\theta_2$, and Figure \ref{fig:pend_res} compares its time evolution computed when using the eABA (implemented in OpenFOAM) and the Lagrangian mechanics (implemented in Python). The Figure show that the angle $\theta_2$ exhibits the same time evolution in both the eABA and the Lagrangian mechanics computation, thereby validating the implementation of eABA. 

\subsubsection{Coupled solver verification}\label{sec_verif}

The coupling of the flow and rigid body dynamic solver is verified using simplified models widely used in literature due to their good aerodynamic predictions for smooth wing kinematics \cite{Sane2002,Lee2016,Cai2021}. The test case is the single-wing drone defined in section \ref{sec:case1}. 

The simplified model is implemented in Python and combines an aerodynamic model with the drone equations of motion. The aerodynamic model is the quasi-steady formulation derived in \cite{Lee2016} which estimates the lift and drag forces along $z_I$ and $x_w\cos \alpha$ (Figure \ref{fig:wing_full}):

\begin{equation} \label{eq:lift}
L_w(t)    = \frac{1}{2}\rho C_{L,w}(\alpha) \int^{R}_{R_0} U_w^2(t,r)  c(r)dr, 
\end{equation}
 \begin{equation} \label{eq:drag}
D_w(t) = \frac{1}{2}\rho C_{D,w}(\alpha) \int^{R}_{R_0} U_w^2(t,r)   c(r)dr, 
\end{equation}
where $C_{L_{w}}$ and $C_{D_{w}}$ are the lift and drag coefficients that have been calibrated with \textit{Fluent} simulations of flapping wings in still air \cite{Lee2016}. Equations \eqref{eq:lift} and \eqref{eq:drag} have been modified from \cite{Lee2016} to make the wing velocity $U_w$ function of the flapping velocity $\dot{\phi}$, the body linear velocity and the body rotational velocity. Validation of the aerodynamic model is found in \cite{Lee2016}.

The drone's equations of motion, derived from aircraft flight dynamics principles \cite{Etkin1995}, describe the translational motion of the drone's center of mass. Specifically, they define the two translational degrees of freedom ($x, z$) (Figure \ref{fig:wing_full}) resulting from the drone's weight ($mg$) and the aerodynamic forces generated by the wings ($F_{x,w}, F_{z,w}$):
 
\begin{align}\label{eq:aircraft}
    m \ddot{x} &= F_{x,w}   \\
    m \ddot{z} &= F_{z,w}  - mg  \label{eq:aircraft_2}
\end{align}
where the wing forces result from the projections of equations \eqref{eq:lift} and \eqref{eq:drag} in the intertial frame. The translational trajectory of the single-wing drone is then obtained from the integration of system \eqref{eq:aircraft}-\eqref{eq:aircraft_2} using a Runge Kutta scheme.  
Implementation details on the simplified environment are given in \cite{Schena2023} and the following of the section compares the two environments using the parameters of Table \ref{tab:parameters}. 
\color{black}

\begin{figure}[!h]
\centering
\includegraphics[width=0.8\textwidth]{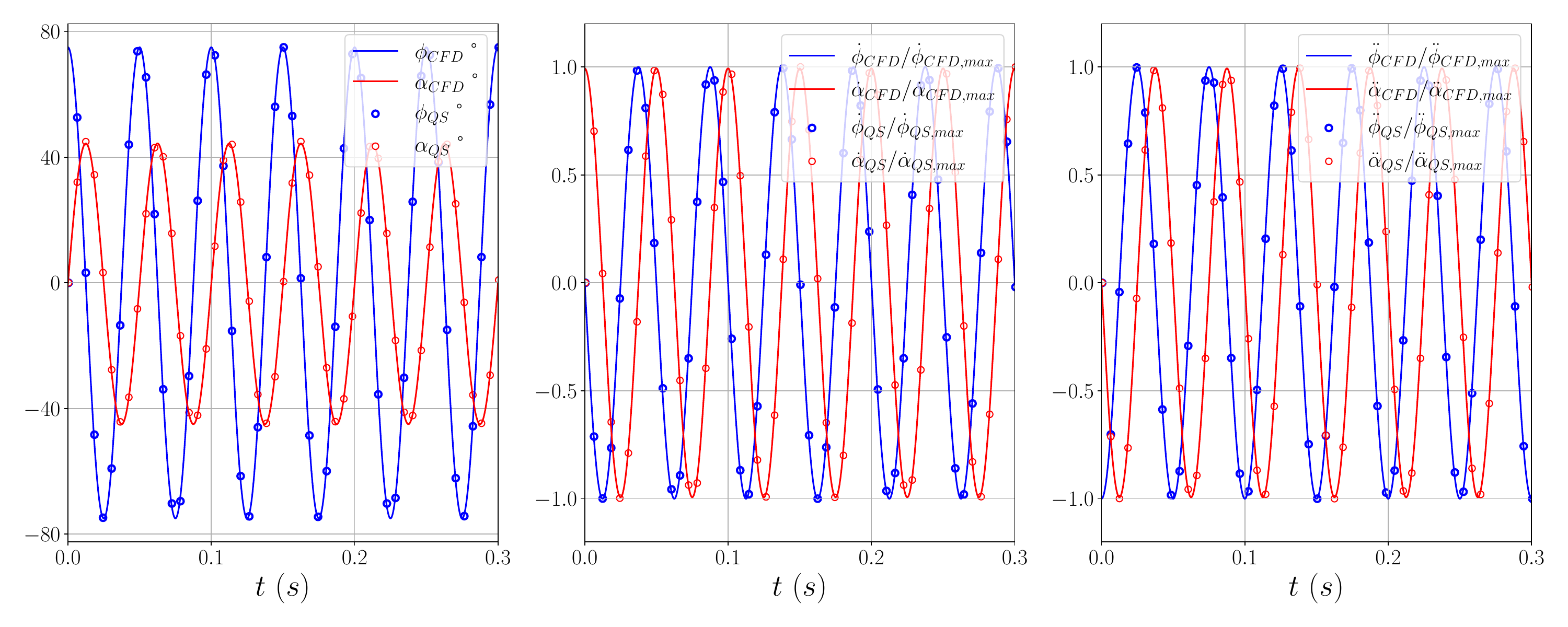}
\caption{Comparison of the flapping and pitching angle (left), velocity (middle), and acceleration (right) between the analytical expression \eqref{eq:phi} and \eqref{eq:alpha} used in the $QS$ model and the motion from the $R_z$ and $R_y$ joints of the multibody drone.}
\label{fig:wingKinematics}
\end{figure}

Figure \ref{fig:wingKinematics} first shows the flapping (blue) and pitching (red) angle, velocity, and acceleration as a function of the travel time for the 6 first cycles.
The figure demonstrates that the ABA algorithm implementation (section \ref{sec:impl}) enables the multibody system's joints to accurately track the kinematics prescribed by equations~\eqref{eq:phi} and~\eqref{eq:alpha}. These equations and their derivatives are directly used in the quasi-steady model (labeled QS).

\begin{figure}[!h]
\centering
\includegraphics[width=0.8\textwidth]{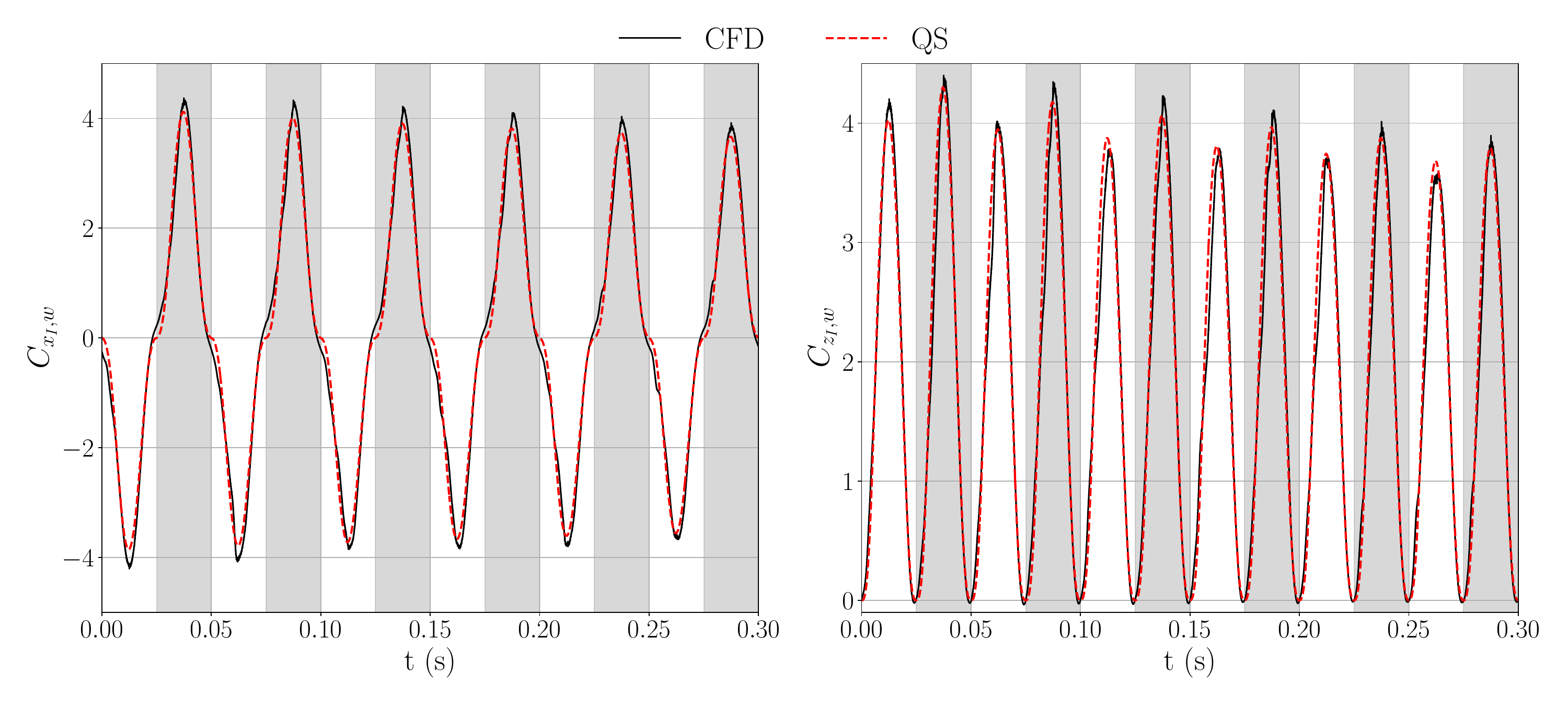}
\caption{Comparison of the wing forces computed in the CFD environment and in the QS model for the longitudinal direction (left) and vertical direction (right) during 20 flapping cycles.}
\label{fig:forcesVerification}
\end{figure}

\color{black}
Figure \ref{fig:forcesVerification} compares the wing forces $C_{x_I,w}$ and $C_{z_I,w}$ (equations \eqref{eq_cxi_w} and \eqref{eq_czi_w}) in the CFD and QS environments.  The first half of the cycle (downstroke) is left unshaded, and the second half (upstroke) is shaded in gray.

The forces from the CFD follow the same trend as the QS forces, verifying the correct implementation of the flow solver, the rigid body dynamics solver and their coupling within the CFD framework.  
Minor discrepancies are primarily attributed to limitations of the simplified model. The QS formulation relies on semi-empirical lift $C_{L_{w}}$ and drag $C_{D_{w}}$ coefficients to model the influence of the LEV. These coefficients were obtained through regression on CFD simulations, which introduces (small) numerical errors.

While the quasi-steady LEV dominates force generation during smooth wing motion, secondary unsteady effects still influence the force profiles to a minor extent \cite{poletti2024}. At stroke reversal, wing–wake interactions introduce small oscillations in the CFD results \cite{Lee2018,Li2022}, further analyzed in Section~\ref{sec:result_2} (see also \cite{poletti2024}). The QS model also omits the Wagner effect, which governs the transient buildup of LEV circulation as the wing accelerates from rest \cite{Van2022}, leading to a slight phase lag between CFD and QS forces. Including such effects would exceed the scope of this verification, particularly given the overall agreement shown in Figure~\ref{fig:forcesVerification}. Under the smooth kinematics considered here, added mass and rotational circulation effects also remain negligible.

\begin{figure}[!h]
\centering
\includegraphics[width=0.8\textwidth]{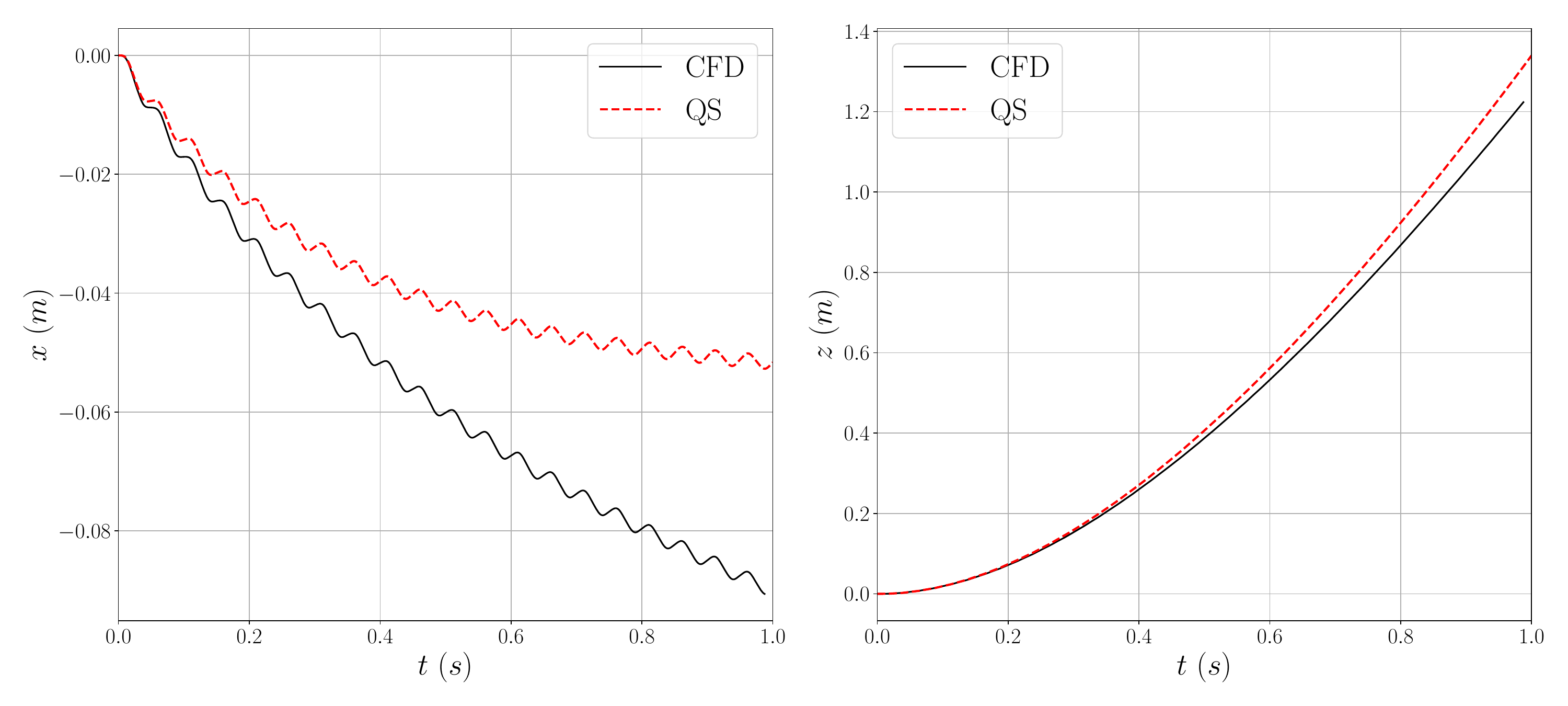}
\caption{Comparison of the drone position computed with the CFD environment and with the QS model for the longitudinal direction (left) and vertical direction (right) during 20 flapping cycles.}
\label{fig:xyzVerification}
\end{figure}

The slightly different wing forces in the CFD and QS lead to minor changes in the drone acceleration. This results in different relative velocities of the wing, which influence the wing force computation, which in turn influences the drone acceleration, etc.
This feedback loop has negligible effects on the drone dynamics, as shown with Figure \ref{fig:xyzVerification}, which presents the complete drone trajectory over time.
The longitudinal ($x$) position shows a maximal discrepancy of 4 cm between the models, while the vertical position ($z$) has a maximal discrepancy of 10 cm.

Interestingly, Figure \ref{fig:xyzVerification} shows that the drone slowly drifts backwards. This is linked to the time evolution of $C_{x_I,w}$. 
During the downstroke (unshaded area), $C_{x_I,w}$ is negative, accelerating the drone backwards. In the upstroke (grey area), $C_{x_I,w}$ is positive, decelerating the drone while the drone is still moving backwards. This explains the oscillations seen on the x-position.
This drift phenomenon strongly depends on the initial state of the wings. 
When the wings start at mid-stroke at $t=0$ s, negative and positive drag contributions cancel each other, resulting in a minimal drone drift. Any residual drift is attributed to the body velocity effect on the wing forces, as also visible in Figure \ref{fig:xyzVerification}. 
At the end of the first cycle, the drone is nearly at rest, but it moves more and more forward from cycle end to cycle end.

\subsection{Computational time characterization}  \label{sec:profiler}
This section analyses the computation time of the simulation described above. The total execution time is divided into three main operations: the rigid body dynamics (RBD), the overset method, and the pimple loop.
The rigid body algorithm starts from the computation of the wing forces and ends with the update of the boundary conditions (Figure \ref{fig:diagram}), the pimple algorithm mainly solves the momentum and pressure equations and the overset method includes all the rest. 

\begin{figure}[!h]
\centering
\includegraphics[width=0.6\textwidth]{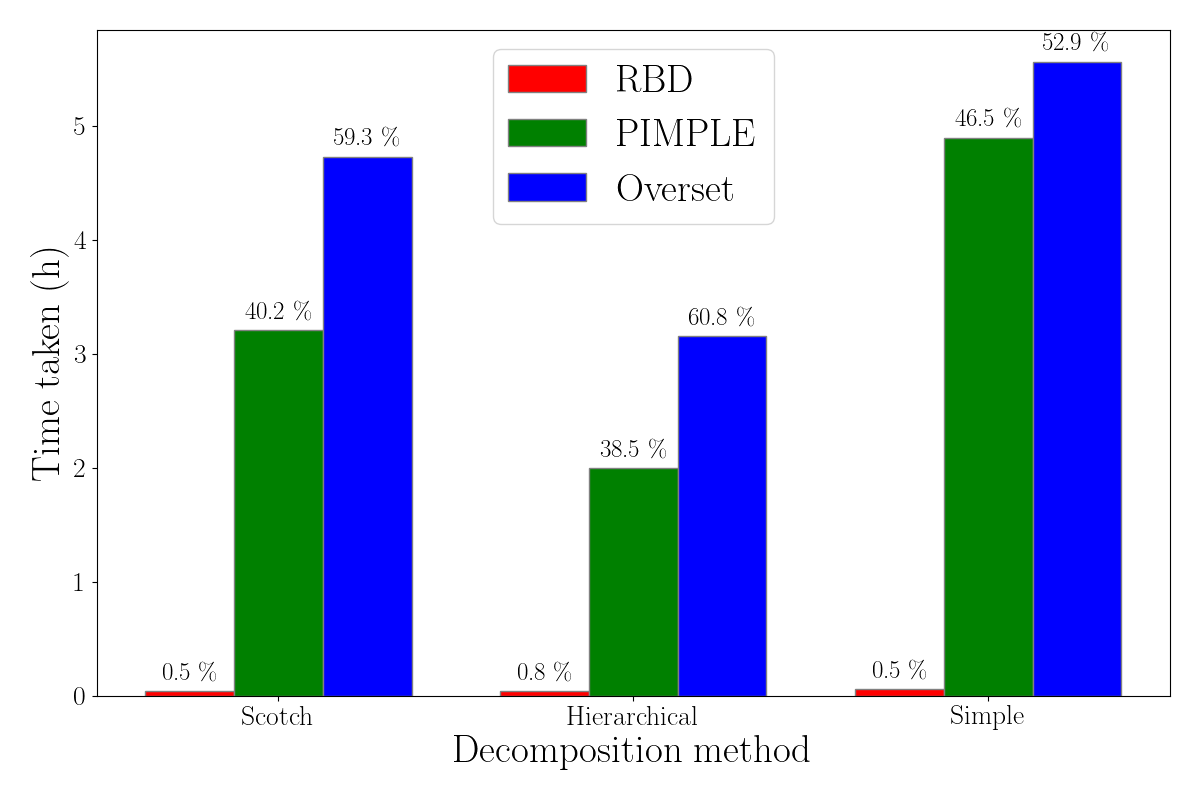}
\caption{Bar chart of the execution time taken by the rigid body dynamics solver, the overset method, and the PIMPLE loop for the scotch, hierarchical, and simple grid decomposition methods.}
\label{fig:pieChart}
\end{figure}

The computations are performed on the 501k cells grid (Figure \ref{fig:xyzGrid}) using 48 processors of the 64 present on one node from the cluster of the Von Karmant Institute. The cluster has 12 nodes with Dual AMD Rome Epyc 7742 and 1TB of RAM in 32 DIMMs for 3200 MHz with 8 memory channels.
Previous studies have shown that the Scotch decomposition results in a high computational cost when using the overset method due to excessive processor communication \cite{Windt2018}. Therefore, three grid decomposition methods, namely Scotch, Hierarchical (zxy), and Simple are tested in Figure \ref{fig:pieChart}. The figure shows the execution time percentages taken by the three operations during 5 flapping cycles. The simulation took around 5 h for the hierarchical decomposition, 8 h for the scotch decomposition, and 10 h for the simple decomposition. 
The execution time overhead added by the rigid body algorithm is negligible, unlike the overset method, which is extremely time demanding ($63\%$ of the total time taken by \textit{overPimpleDyMFoam} with the scotch method). The overset method calls three main functions\cite{ofWiki}: an \texttt{update} function that updates the roles of the cells according to the wing motion (\textit{inverseDistanceCellCellStencil.C}), a \texttt{updateAdressing}  function that excludes holes and includes interpolated cells in the equation matrices (\textit{dynamicOversetFvMesh.C}), and an \texttt{interpolateFields}  function in \textit{cellCellStencilTemplates.C} that interpolate the fields from the component grid to the background grid and vice versa. 
The first operation involves updating the position of the hole cells and interpolated cells as well as searching for donor cells from which interpolation is performed. This last operation is especially time-consuming. Possible improvements are explored in \cite{oversetImpl}.


\section{Results from the body and wing drone}\label{sec:result_2}

This section analyses the flight performances of the BW drone described in section \ref{sec:case2}.

\subsection{Longitudinal motion} \label{sec_long}

\begin{figure}[!h]
\centering
\includegraphics[width=0.8\textwidth]{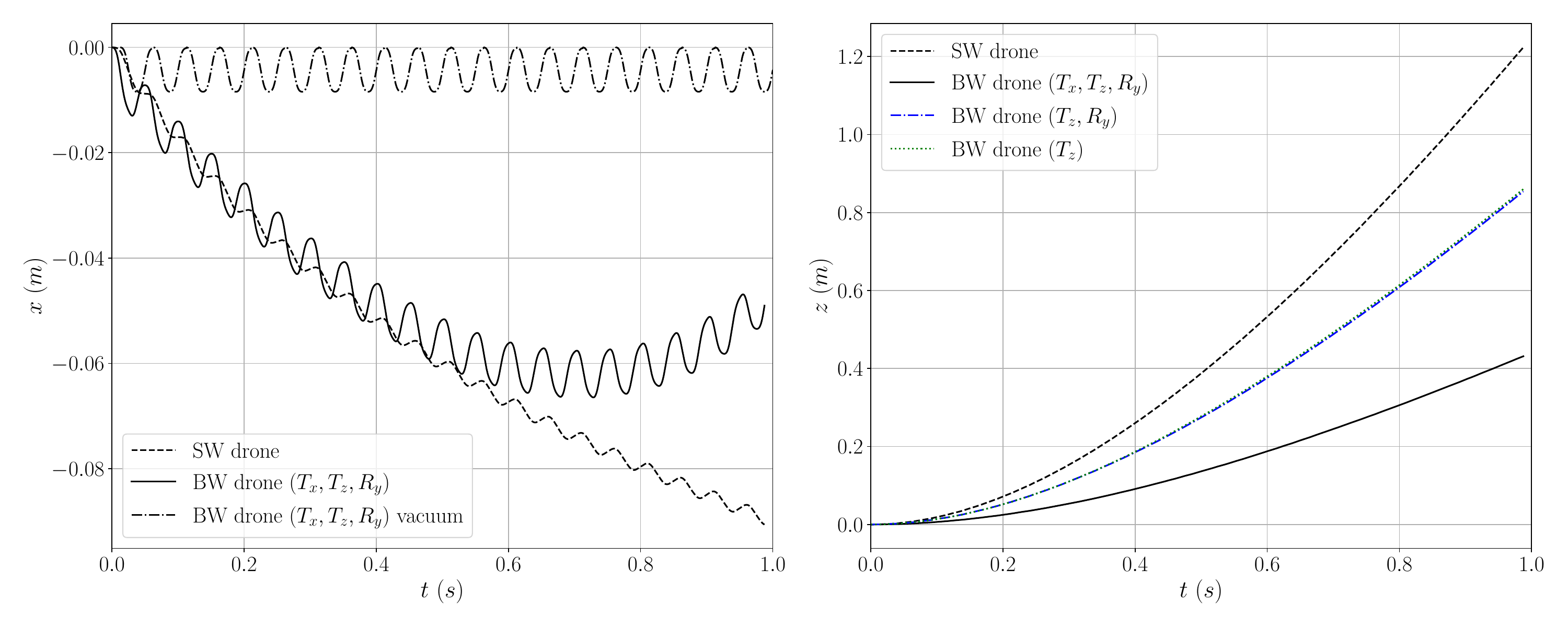}
\caption{Horizontal (left) and vertical (right) position of the drone body as a function of time for 20 flapping periods.}
\label{fig:xyz_body}
\end{figure}

Figure \ref{fig:xyz_body} (left) shows the time evolution of the horizontal position of the drone's centre of mass. The trajectory followed by the SW drone is also shown as a dashed line. Both drones experience backwards drift due to the influence of wing drag (as discussed in the previous section). The BW drone exhibits larger peak-to-peak oscillations due to its inertial coupling with the wing: as the wing flaps in one direction (e.g., toward $x<0$), the body moves slightly in the opposite direction (e.g., toward $x>0$). 
To isolate this effect, Figure \ref{fig:xyz_body} (left) also shows the trajectory of the BW drone flying in a vacuum, where no aerodynamic forces are present (Equation \eqref{eq_fe}). These kinematically induced oscillations combine with the aerodynamically induced oscillations detailed in Section \ref{sec_verif}. \textcolor{black}{One can also notice that at $t=0.7$ s, the drone starts moving in the opposite direction. This is explained by the larger pitching angle $\theta$ as detailed in Section \ref{sec:flow}}

\subsection{Vertical motion}
The SW and BW drones generate an average lift exceeding their weight, causing them to ascend (Figure \ref{fig:xyz_body} right).
However, the BW drone reaches a height $\sim0.7$ m lower than the SW drone.
To illustrate this difference, Figure \ref{fig:xyz_body} (right) presents two additional trajectories obtained by simulating the drones with a smaller degree of freedom set. The figure shows (1) a BW drone with a body featuring two degrees of freedom $(T_z,R_y)$, represented by the dash-dot blue line, and (2) a BW drone with a single degree of freedom body $(T_z)$, shown as a dotted green line.

\begin{figure}[!h]
\centering
\includegraphics[width=0.8\textwidth]{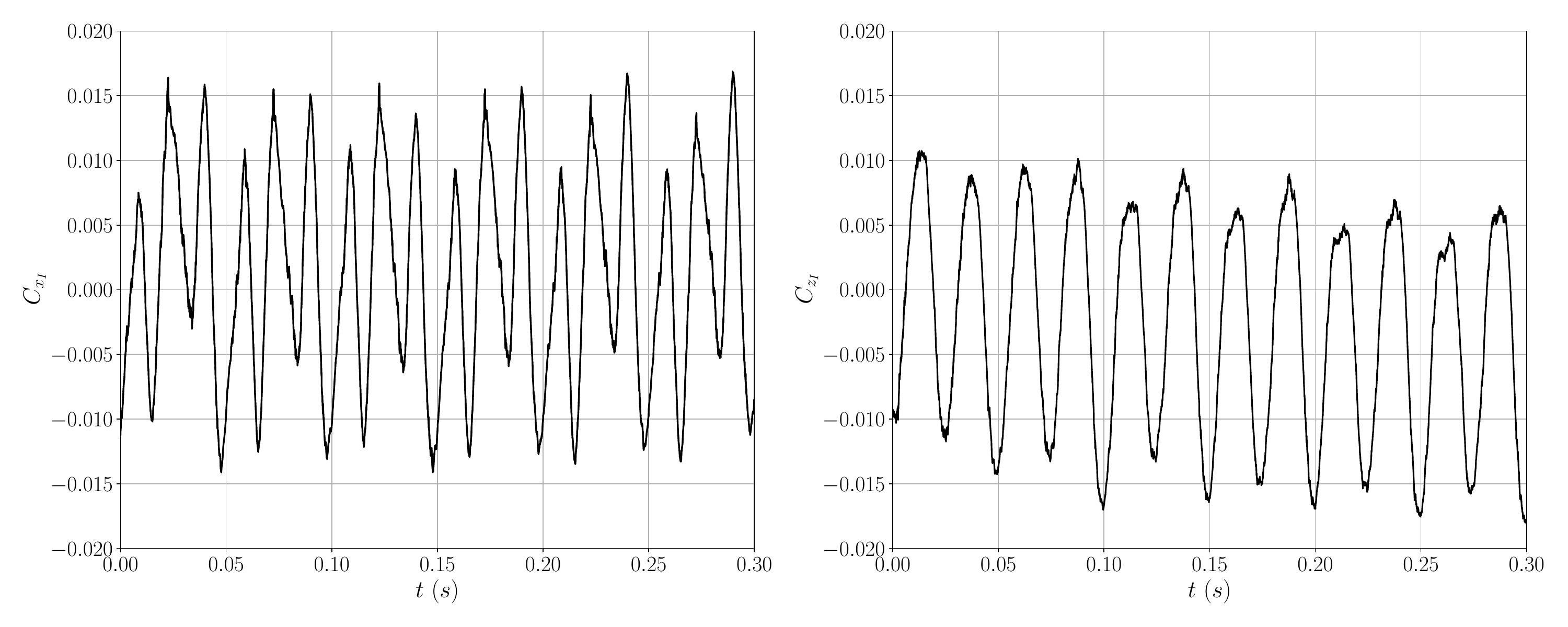}
\caption{Horizontal (left) and vertical (right) forces of the drone body as a function of time for the 6 first flapping periods.}
\label{fig:force_body}
\end{figure}

Comparing these trajectories reveals that the longitudinal oscillations of the full-DOF BW drone tend to slow its ascent. As the body moves opposite to the wing, the relative airflow over the wing decreases, reducing the vertical force $F_z$. Additionally, the body contributes a vertical drag force opposing the motion (Figure~\ref{fig:force_body}, right), which becomes increasingly negative as the drone climbs. However, its impact is minor, being two orders of magnitude smaller than the wing-induced drag.

The BW drone also slightly pitches up and down which tilts the stroke plane, slightly increasing $F_x$ and decreasing $F_z$. These small pitching angles do not explain the large z-position discrepancies, as confirmed by comparing the trajectories the one dof body ($T_z$) and two dofs body $(T_z, R_y)$. This differs from observations made in \cite{Wu2009} where larger pitching motions were reported to strongly change insect trajectories.


\subsection{Pitching dynamics} 
\begin{figure*}[!ht]
    \centering

    \begin{subfigure}{0.9\textwidth}
        \centering
        \includegraphics[width=\textwidth]{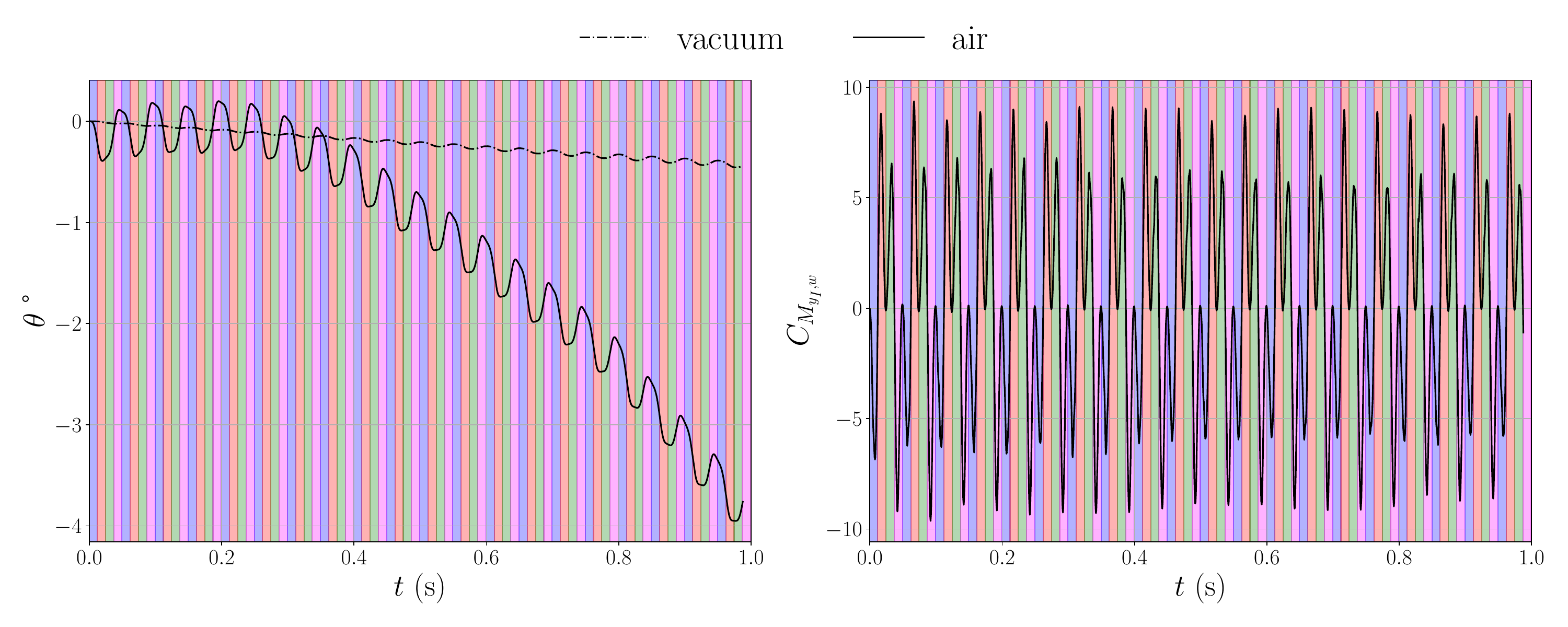}
        \caption{}
        \label{fig:pitching_body}
    \end{subfigure}

    \hfill

    \begin{subfigure}{0.4\textwidth}
        \centering
        \includegraphics[width=\textwidth]{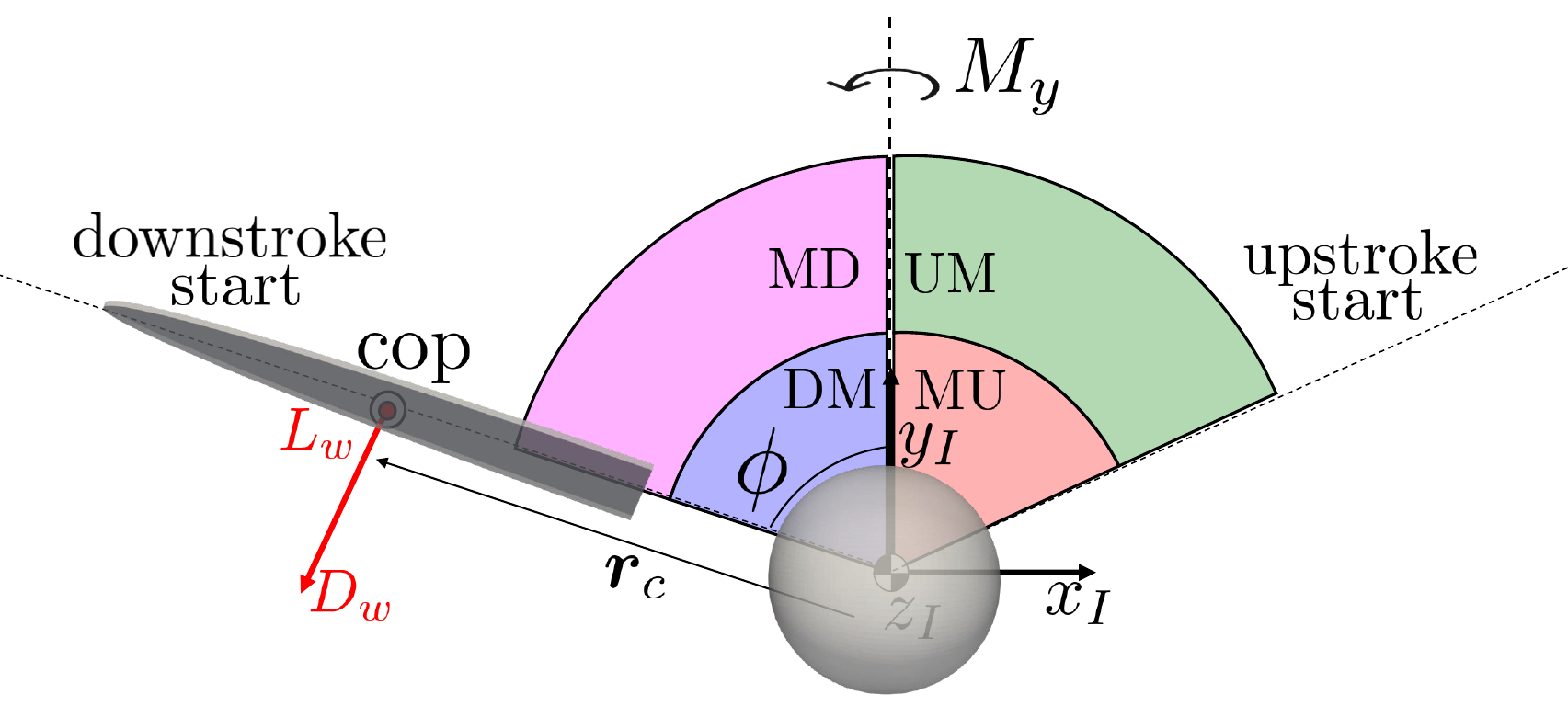}
        \caption{}
        \label{fig:scMy}
    \end{subfigure}

    \caption{\textcolor{black}{Body pitching dynamics shown with (a) the body pitching angle, (b) the wing pitching torque, and (c) a schematic of the drone and the wing forces.}}
    \label{fig:pitching_body_full}
\end{figure*}

\color{black}
Figure \ref{fig:pitching_body} shows the body pitch angle $\theta$ (left) and the pitching torques $C_{M_{y_I,w}}$ induced by wing forces to the centre of mass of the body (right). 
The cycle is divided into four distinct phases, illustrated by different colours in Figures \ref{fig:pitching_body} and \ref{fig:scMy}: downstroke start to midstroke (DM), midstroke to upstroke start (MU), upstroke start to midstroke (UM), and midstroke to downstroke start (MD).
\color{black}
The body starts with $\theta = 0^{\circ}$, indicating that the stroke plane is parallel to the ground (Figure \ref{fig:wing_full}).
During the downstroke (DM and MU), $\theta$ predominantly decreases, while it increases during the upstroke (UM and MD).
The first cycle ends with $\theta > 0^{\circ}$, indicating a nose-down body orientation and a tilted stroke plane. 
\color{black}
This small deviation from $\theta = 0^{\circ}$ carries over to subsequent cycles, after which the pitching angle gradually decreases throughout the simulation. 
The body orientation is therefore passively unstable, which was also observed in various flapping species \cite{Schenato2003}.
This instability arises primarily from the wing kinematics and aerodynamic forces. 
The wing kinematics influences the body pitching dynamics analogously to its influence on the longitudinal dynamics discussed in Section \ref{sec_long}.
Specifically, the wing pitching, combined with the flapping motion, induces a counter-pitching motion of the body due to a reaction torque at the body-wing joints. This torque gradually reduces the body's pitching angle over successive flapping cycles as shown in Figure \ref{fig:pitching_body} (left) with the dash-dot line showing the trajectory of the BW drone simulated in a vacuum. 

Nevertheless, the influence of the wing kinematics on the pitching dynamics is surpassed by the aerodynamic excitation of the wings. The latter is less trivial to understand. A first analysis could conclude in torque equilibrium as the lift should generate a torque during DM (UM) that is counter-balanced by the torque during MU (MD). Figure \ref{fig:pitching_body} (right) shows that the pitching torque has indeed four torque peaks during each part of the cycles but the peaks from DM and UM are smaller ($|C_{M_{y_I,w}}| \sim 4$) than the ones from MU and MD ($|C_{M_{y_I,w}}| \sim 6$). 
This asymmetry has three main origins, explained with the help of a schematic in Figure \ref{fig:scMy} and the constitutive equation of the pitching moment:
\begin{equation}\label{eq:My}
    M_y = r_{cz} F_x - r_{cx} F_z,
\end{equation}
where $\bm{r_c}=[r_{cx},r_{cy},r_{cz}]$ is the lever arm, i.e., the distance between the center of pressure and the center of rotation (which is set to the center of mass). 
\color{black}
Firstly, the forces $F_x,F_z$ are not equal during the downstroke (DM and MU) and upstroke (UM and MD) because they are functions of the body velocity. For example, the LEV contribution is a function of the relative wing-air velocity, which is itself a function of the body velocity. The wing-wake interaction contribution also varies with the drone motion. At the beginning of the flight, when the drone is almost hovering, the flow is not fully developed, and the wing faces different wake flows during the successive downstrokes and upstrokes. 
Afterwards, the wing encounters fewer wake structures because it continuously accelerates upwards. 

Secondly, the amplitude of the forces is not symmetric within a stroke due to the Wagner effect. The time integral of the lift during the second stroke halves (MU and MD) is larger than during the first halves. Consequently, the pitching moment is increased during the second half of the strokes.

\color{black}
Thirdly, even if one ignores the previously described asymmetry sources, the pitching moment would still give two different amplitude peaks due to the center of pressure that is offset from the wing symmetry axis (i.e. $r_{cz}\neq0$). 

To clearly explain this, one can look at the different signs from the two contributions of equation \eqref{eq:My} during a flapping cycle.
Looking at Figure \ref{fig:scMy}, knowing that $r_{cz}$ is negative, one can assume that the first contribution $r_{cz} F_x$ is positive during DM (negative during UM). The drag keeps its sign during a stroke, and so does $r_{cz} F_x$ during the second half of the stroke. 
On the other hand, for the second contribution, $r_{cx}$ changes sign during one stroke (Figure \ref{fig:scMy}), inducing a nose-up and nose-down motion during each half-stroke. 
The summation of the two terms leads to an increase or decrease of the pitching moment amplitude.

\color{black}


\subsection{Flowfield analysis} \label{sec:flow}
\begin{figure}[!h]
\centering
\includegraphics[width=0.5\textwidth]{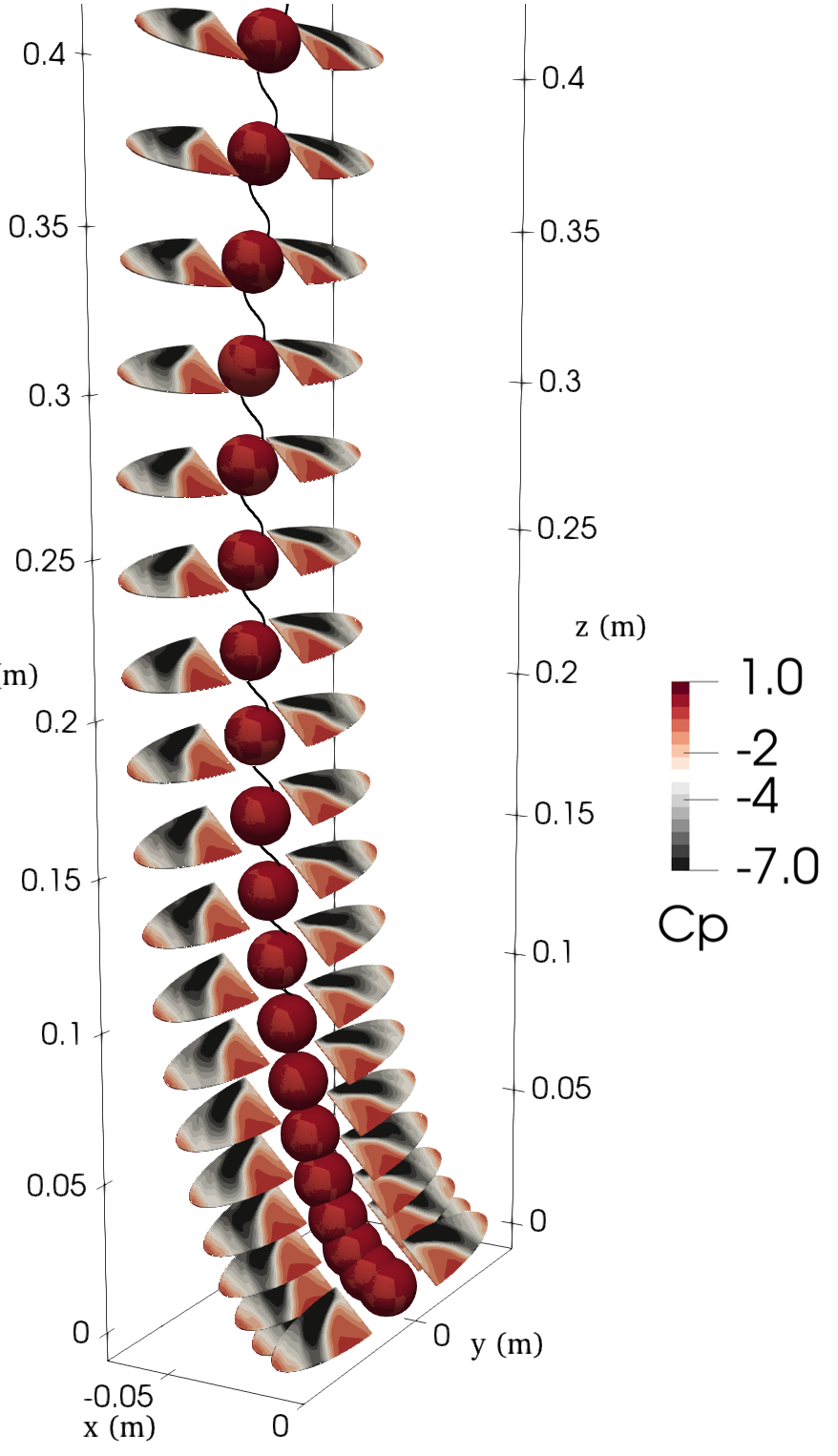}
\caption{Pressure coefficient on the body-wing drone shown at midpoints of the upstrokes during its vertical ascent trajectory}
\label{fig:field1}
\end{figure}

\color{black}
This section illustrates the aerodynamic phenomena driving the drone's motion through the time evolution of the pressure field, velocity field and vortical structures.
Figure \ref{fig:field1} shows the $x-z$ trajectory overlaid with snapshots of the drone at the midpoints of the upstrokes across successive flapping cycles. As explained in Section \ref{sec:case1}, the drone was simulated with only one wing and the second wing shown in Figure \ref{fig:MBS2} is its mirror image across the $(x,z)_I$ symmetry plane.

The figure first evidences the longitudinal oscillations of the body due to the wing-body inertial coupling and aerodynamic forces. While the tilting of the stroke plane is too small to be detected, its effects are seen on the $x-z$ trajectory. The drone initially moves towards the $x<0$ ($z\sim\in[0,0.1]$) due to the wing drag as discussed in Section \ref{sec_long}. As the magnitude of the pitching angle and thus the stroke plane angle increases (see Figure \ref{fig:pitching_body_full}), the drone moves in the opposite direction due to the redirection of the lift force in the horizontal direction.

\begin{figure}[!h]
\centering
\includegraphics[width=0.7\textwidth]{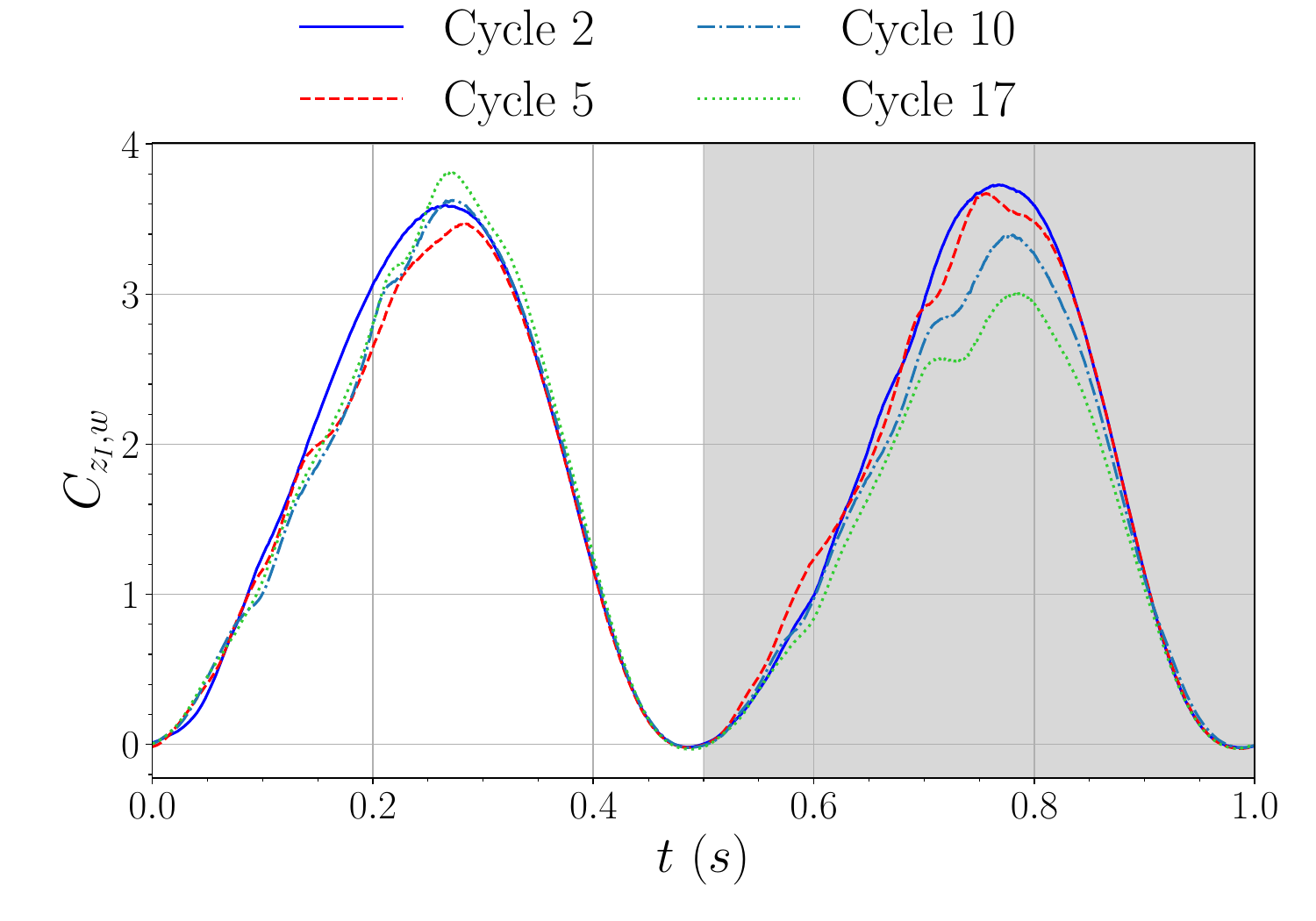}
\caption{Lift force during the second, fifth, tenth, and seventeenth flapping cycles.}
\label{fig:fzPitch}
\end{figure}

This effect is well captured by Figure \ref{fig:fzPitch}, which shows the vertical force during the second, fifth, tenth, and seventeenth flapping cycles. The forces are similar during the second and fifth cycles, with the fifth cycle showing more lift oscillations due to the wing-wake interaction effect arising when the flow is developing. However, from the fifth to the seventeenth cycle, the peak force increases in the downstroke and decreases in the upstroke. This results from the body pitching motion with time, setting the stroke plane more vertical and so increasing the angle of attack during the downstrokes and decreasing it during the upstrokes. One can notice that this effect is stronger during the upstrokes.

\begin{figure*}[!ht]
    \centering

    \begin{subfigure}{0.7\textwidth}
        \centering
        \includegraphics[width=\textwidth]{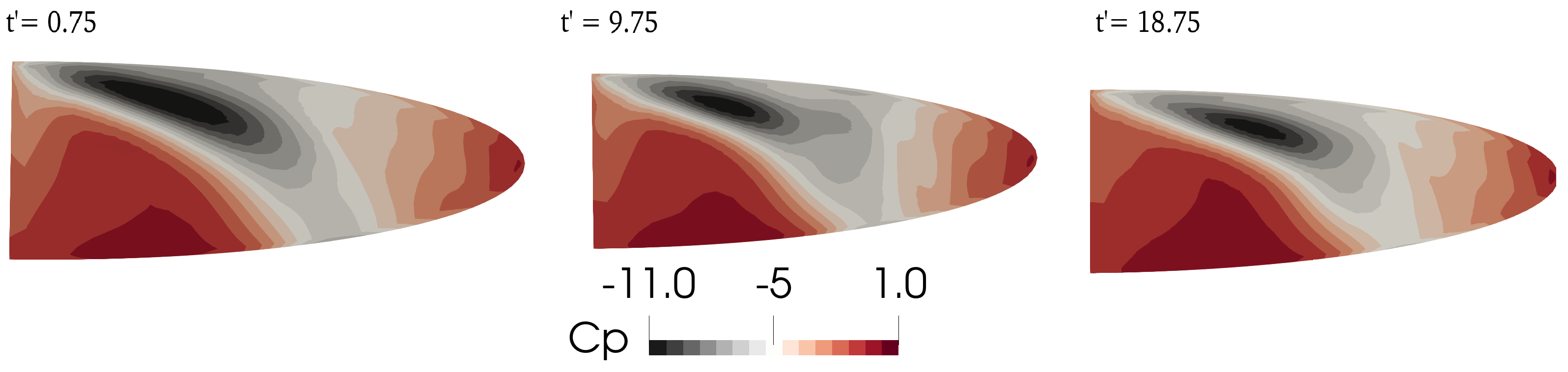}
        \caption{}
        \label{fig:pSuct}
    \end{subfigure}

    \begin{subfigure}{0.7\textwidth}
        \centering
        \includegraphics[width=\textwidth]{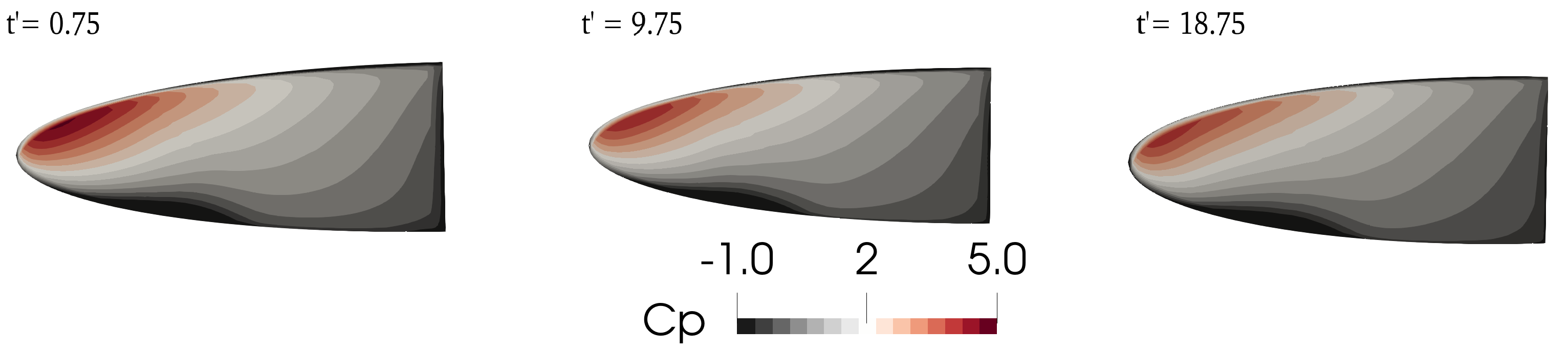}
        \caption{}
        \label{fig:pPres}
    \end{subfigure}

        \begin{subfigure}{0.7\textwidth}
        \centering
        \includegraphics[width=\textwidth]{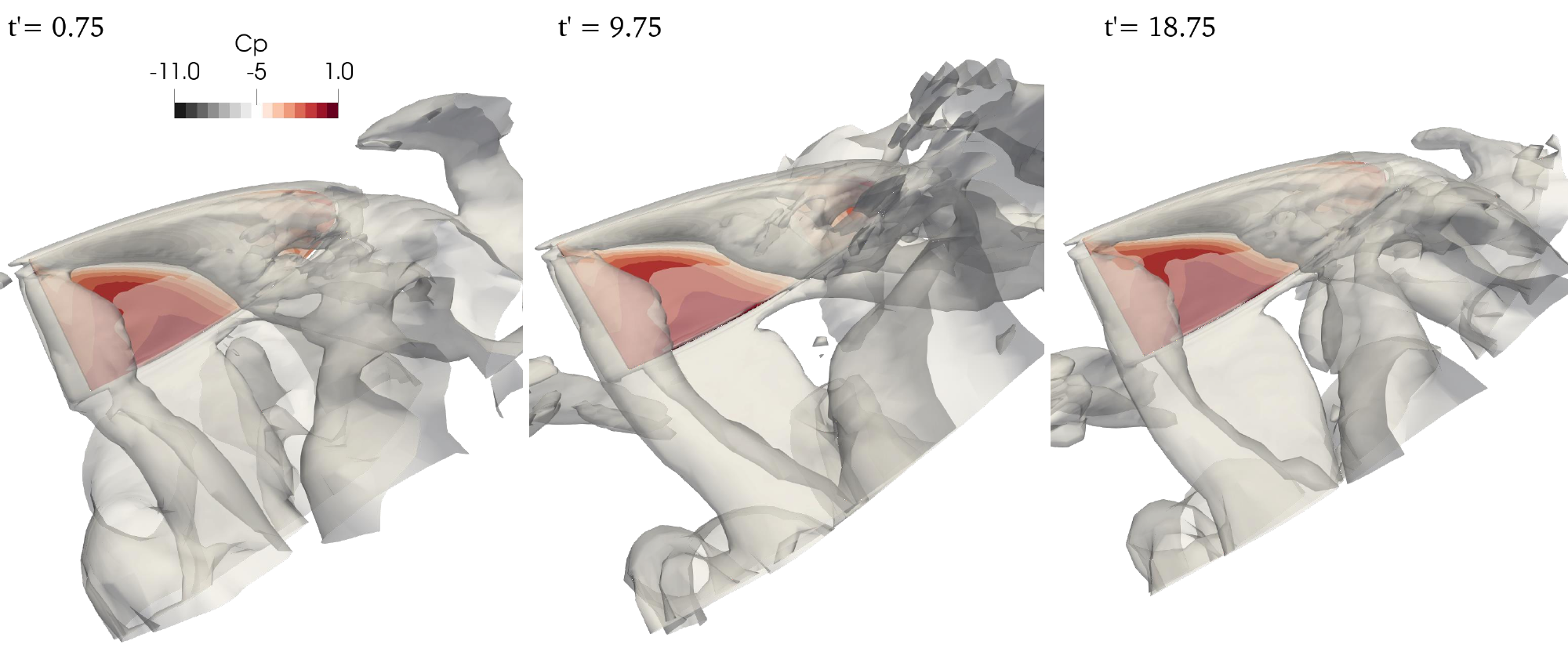}
        \caption{}
        \label{fig:lev}
    \end{subfigure}
    \caption{\textcolor{black}{Pressure distribution on (a) the suction side and (b) the pressure side of the wing at midpoints of the upstrokes for the first, ninth and eighteenth cycle. The LEV is shown in (c) using the isocontours $Q>0$ and $\omega_{y_w}<0$}.}
    \label{fig:levAll}
\end{figure*}

Figure \ref{fig:field1} also shows the pressure distribution on the drone's surface. The wings experience the strongest (negative) pressure on their suction side,  confirming that this side drives the drone upwards. During the flight, the mean pressure slightly decreases while the pressure distributions remain similar. 
Figure \ref{fig:levAll} focuses on midpoints of the upstrokes corresponding to $t'=[0.75,9.75,18.75]$.  Figure \ref{fig:pSuct} shows the pressure distribution on the suction side, Figure \ref{fig:pPres} on the pressure side and Figure \ref{fig:lev} relates the pressure distribution to the LEV using iso-contours of the Q-criterion ($Q>0$) \cite{Jeong1995} and the spanwise vorticity ($\omega_{y_w}>0$). The pressure distributions on the pressure side are almost identical, with a larger pressure near the tip due to the highest wing velocity. On the suction side, the low-pressure region forms a triangular shape which is the fingerprint of the leading-edge vortex (LEV) as seen in Figure \ref{fig:lev}. The LEV is stably attached to the wing and grows along the span due to its dependence on the flapping velocity. 
The LEV core pressure decreases as the drone ascends due to the decrease of the effective angle of attack and the increase of the pitch angle magnitude. Table \ref{tab:lev} confirms this trend with the LEV volume $V'$ and circulation $\Gamma$ computed as: 

\begin{equation}
    \Gamma = \int_{V'} \omega_{y_w} dV'\quad V':Q>0 \quad \cap \quad \omega_{y_w}>0
\end{equation}
\color{black}

\begin{figure}[!h]
\includegraphics[width=\textwidth]{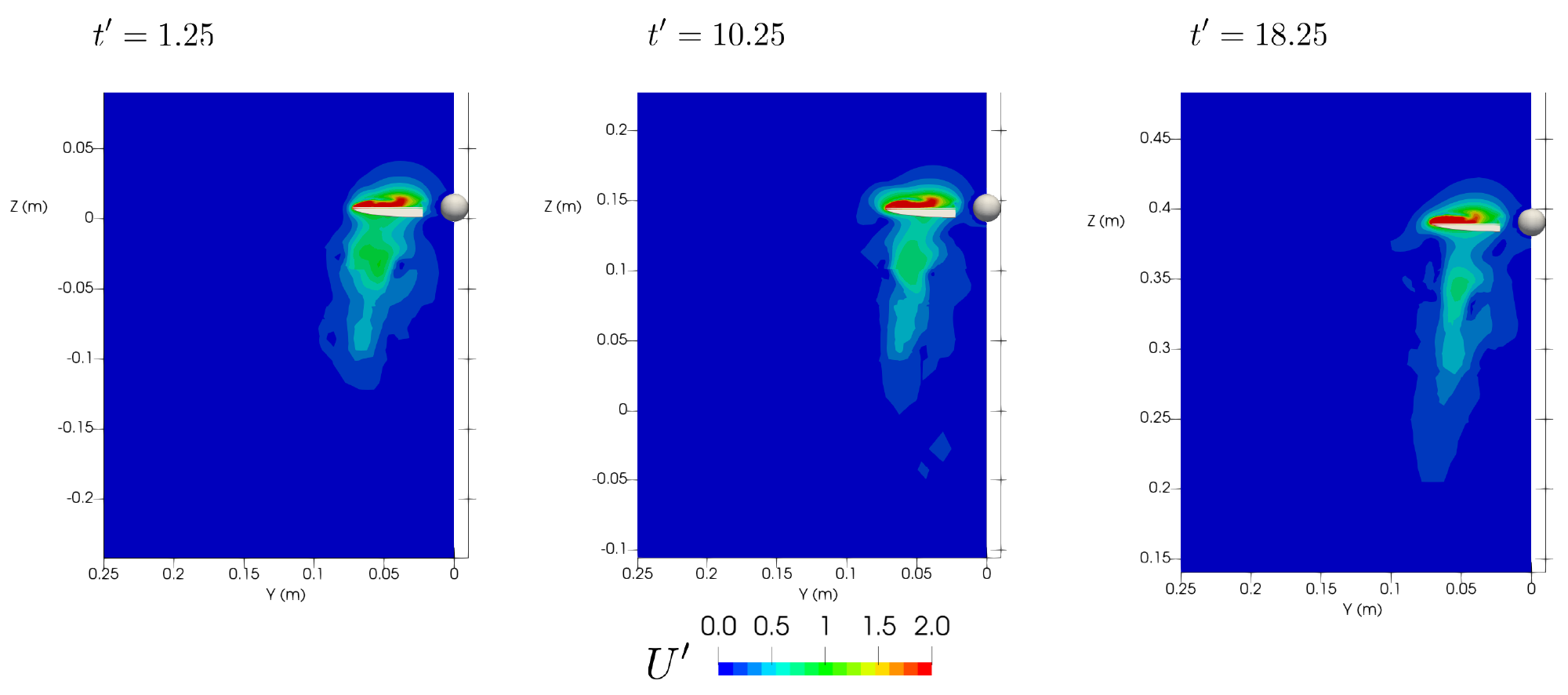}
\caption{Wake formation evidenced by the dimensionless velocity $U'=U/U_{ref}$ field during the drone vertical motion for three time steps.}
\label{fig:field3}
\end{figure}

Finally, Figure \ref{fig:field3} shows three clipped slices of the moving domain that are symmetry planes of the drone taken at the midstroke of the second,  eleventh, and nineteenth cycles. The wake develops along the vertical direction and two patterns corresponding to the current and previous cycle can be identified. The previous wake structures are smeared out in the domain due to the largest cells used further away from the drone. 

\begin{figure}[!h]
\includegraphics[width=\textwidth]{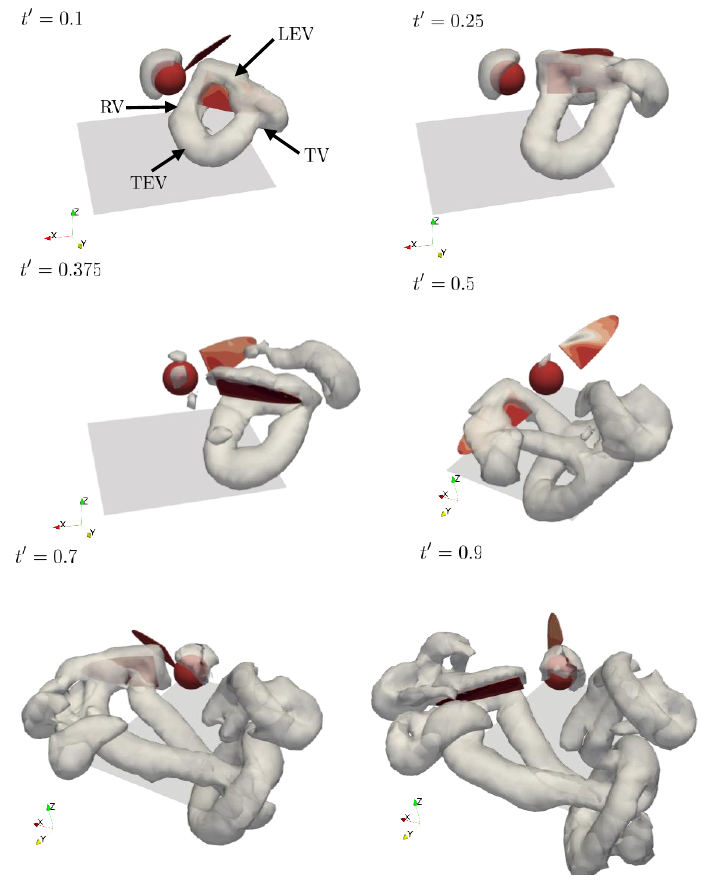}
\caption{\textcolor{black}{Vortex structures generated by the body-wing drone during the first cycle. LEV: leading-edge vortex, TEV: trailing-edge vortex, TV: tip vortex, RV: root vortex.}}
\label{fig:field4}
\end{figure}

\textcolor{black}{
Figure \ref{fig:field4} shows the evolution of the vortex structures during the first cycle.
The structures appear thicker than those shown in Figure \ref{fig:lev} as the contours of the Q-criterion were computed on the background grid to show the full length of the structures. In contrast to the previous analyses, the mid-position of the upstroke is taken as the beginning ($t'=0$). 
Shortly after the start, the wings show three clear clockwise vortex structures forming a loop. The leading-edge vortex (LEV) sheds parts of its vorticity through a tip vortex (TV), and the body-wing clearance allows the development of a root vortex (RV). The trailing-edge vortex, shed shortly after the wing starts in the wake, connects these three structures.
At stroke reversal, the wing decreases its pitching angle (i.e. wing becomes more vertical) and the vortex loop detaches from the wing. For the next stroke, the wing accelerates in the other direction and generates a new attached LEV, TV and RV forming a second vortex loop closed by the previously shed vortices.}

\begin{table}[!ht]
\centering
\begin{tabular}{c|c|c|c}
\hline
  &  $t'=0.75$  & $t'=9.75$ & $t'=18.75$ \\ \hline  
$V/V_1$ $(\%)$ &  1  &  99 &  92  \\
$\Gamma/\Gamma_1$ $(\%)$&  1  &  92 & 90 \\
\hline
\end{tabular}
\caption{Comparison of the LEV size and strength for $t'=[0.75,9.75,18.75]$ }
\label{tab:lev}
\end{table}


\section{Conclusion}

This work characterizes and extends the rigid body dynamics solver in OpenFOAM v2206 to enable high-fidelity simulations of flapping-wing drones. The solver is based on the Articulated-Body Algorithm (ABA), which integrates the equations of motion for multibody tree systems such as those found in flapping-wing configurations. To allow a subset of joints to follow prescribed motions while others evolve freely, we introduced active articulations by extending the ABA and modifying the governing equations of motion.

These developments are encapsulated in a new \textit{imposedMotion} library, which supports hybrid multibody simulations with both imposed and free motions—capabilities that were previously unavailable in OpenFOAM. While motivated by flapping-wing drones, the framework is broadly applicable to other multibody systems, including biological flyers, propeller-driven vehicles, and marine platforms.

The framework was first verified using a double pendulum benchmark derived from Lagrangian mechanics. It was then applied to two drone configurations: a single-wing drone and a body-wing drone with up to five degrees of freedom. Simulations of ascending flight were performed using the \textit{overPimpleDyMFoam} solver with overset meshes, a moving background grid, and an LES turbulence model. Drone trajectories showed good agreement with quasi-steady predictions, and parametric studies confirmed the numerical robustness of the solver. A detailed profiling analysis identified the overset grid update as the main computational bottleneck.

Flow visualizations and force analyses revealed strong oscillatory behavior in the body-wing drone, driven by the coupling between wing kinematics and body inertia. These results highlight the importance of resolving both aerodynamic and inertial effects in free-flight simulations.

Future work will focus on (1) coupling the CFD framework with a structural solver from Kratos Multiphysics via the CoCoNuT code \cite{kratos,coconut}, and (2) simulating dynamic maneuvers using wing kinematics optimized for various objectives with simplified models.

With these enhancements, the proposed simulation framework could serve as a digital companion to real flapping-wing drones—bridging the gap between biological flight performance and current engineering designs, which have so far relied on limited CFD modeling.

\section*{Acknowledgment}
Romain Poletti is supported by Fonds Weten-
schappelijk Onderzoek (FWO), Project No. 1SD7823N. 







\bibliographystyle{IEEEtran}

\bibliography{cas-refs}

\begin{thebibliography}{10}
\providecommand{\url}[1]{#1}
\csname url@samestyle\endcsname
\providecommand{\newblock}{\relax}
\providecommand{\bibinfo}[2]{#2}
\providecommand{\BIBentrySTDinterwordspacing}{\spaceskip=0pt\relax}
\providecommand{\BIBentryALTinterwordstretchfactor}{4}
\providecommand{\BIBentryALTinterwordspacing}{\spaceskip=\fontdimen2\font plus
\BIBentryALTinterwordstretchfactor\fontdimen3\font minus \fontdimen4\font\relax}
\providecommand{\BIBforeignlanguage}[2]{{%
\expandafter\ifx\csname l@#1\endcsname\relax
\typeout{** WARNING: IEEEtran.bst: No hyphenation pattern has been}%
\typeout{** loaded for the language `#1'. Using the pattern for}%
\typeout{** the default language instead.}%
\else
\language=\csname l@#1\endcsname
\fi
#2}}
\providecommand{\BIBdecl}{\relax}
\BIBdecl

\bibitem{Cheng2016}
B.~Cheng, B.~W. Tobalske, D.~R. Powers, T.~L. Hedrick, S.~M. Wethington, G.~T. Chiu, and X.~Deng, ``Flight mechanics and control of escape manoeuvres in hummingbirds. i. flight kinematics,'' \emph{Journal of Experimental Biology}, vol. 219, no.~22, pp. 3518--3531, 2016.

\bibitem{Taylor2002}
G.~Taylor and A.~Thomas, ``Animal flight dynamics ii. longitudinal stability in flapping flight,'' \emph{Journal of theoretical biology}, vol. 214, no.~3, pp. 351--370, 2002.

\bibitem{Karasek2012}
M.~Kar{\'a}sek and A.~Preumont, ``Flapping flight stability in hover: A comparison of various aerodynamic models,'' \emph{International Journal of Micro Air Vehicles}, vol.~4, no.~3, pp. 203--226, 2012.

\bibitem{Schenato2003}
L.~Schenato, D.~Campolo, and S.~Sastry, ``Controllability issues in flapping flight for biomimetic micro aerial vehicles (mavs),'' in \emph{42nd IEEE International Conference on Decision and Control (IEEE Cat. no. 03CH37475)}, vol.~6.\hskip 1em plus 0.5em minus 0.4em\relax IEEE, 2003, pp. 6441--6447.

\bibitem{Deng2006}
X.~Deng, L.~Schenato, and S.~S. Sastry, ``Flapping flight for biomimetic robotic insects: Part ii-flight control design,'' \emph{IEEE transactions on robotics}, vol.~22, no.~4, pp. 789--803, 2006.

\bibitem{Chen2020}
L.~Chen, J.~Wu, and B.~Cheng, ``Leading-edge vortex formation and transient lift generation on a revolving wing at low reynolds number,'' \emph{Aerospace Science and Technology}, vol.~97, p. 105589, 2020.

\bibitem{Meng2015}
X.~G. Meng and M.~Sun, ``Aerodynamics and vortical structures in hovering fruitflies,'' \emph{Physics of Fluids}, vol.~27, no.~3, 2015.

\bibitem{Liu2020}
L.-G. Liu, G.~Du, and M.~Sun, ``Aerodynamic-force production mechanisms in hovering mosquitoes,'' \emph{Journal of Fluid Mechanics}, vol. 898, p. A19, 2020.

\bibitem{Lee2018}
Y.~Lee and K.~B. Lua, ``Wing--wake interaction: comparison of 2{D} and 3{D} flapping wings in hover flight,'' \emph{Bioinspiration \& Biomimetics}, vol.~13, no.~6, p. 066003, 2018.

\bibitem{Song2014}
J.~Song, H.~Luo, and T.~L. Hedrick, ``Three-dimensional flow and lift characteristics of a hovering ruby-throated hummingbird,'' \emph{Journal of The Royal Society Interface}, vol.~11, no.~98, p. 20140541, 2014.

\bibitem{Aono2008}
H.~Aono, F.~Liang, and H.~Liu, ``Near-and far-field aerodynamics in insect hovering flight: an integrated computational study,'' \emph{Journal of Experimental Biology}, vol. 211, no.~2, pp. 239--257, 2008.

\bibitem{Liu2021}
Y.~Liu, A.~D. Lozano, T.~L. Hedrick, and C.~Li, ``Comparison of experimental and numerical studies on the flow structures of hovering hawkmoths,'' \emph{Journal of Fluids and Structures}, vol. 107, p. 103405, 2021.

\bibitem{Bomphrey2017}
R.~J. Bomphrey, T.~Nakata, N.~Phillips, and S.~M. Walker, ``Smart wing rotation and trailing-edge vortices enable high frequency mosquito flight,'' \emph{Nature}, vol. 544, no. 7648, pp. 92--95, 2017.

\bibitem{Cai2021}
X.~Cai, D.~Kolomenskiy, T.~Nakata, and H.~Liu, ``A {CFD} data-driven aerodynamic model for fast and precise prediction of flapping aerodynamics in various flight velocities,'' \emph{Journal of Fluid Mechanics}, vol. 915, p. A114, 2021.

\bibitem{Luo2012}
H.~Luo, H.~Dai, P.~J.~F. de~Sousa, and B.~Yin, ``On the numerical oscillation of the direct-forcing immersed-boundary method for moving boundaries,'' \emph{Computers \& Fluids}, vol.~56, pp. 61--76, 2012.

\bibitem{Zheng2013}
L.~Zheng, T.~Hedrick, and R.~Mittal, ``A comparative study of the hovering efficiency of flapping and revolving wings,'' \emph{Bioinspiration \& biomimetics}, vol.~8, no.~3, p. 036001, 2013.

\bibitem{Bos2013_2}
F.~M. Bos, B.~W. van Oudheusden, and H.~Bijl, ``Radial basis function based mesh deformation applied to simulation of flow around flapping wings,'' \emph{Computers \& Fluids}, vol.~79, pp. 167--177, 2013.

\bibitem{Deng2016}
S.~Deng, T.~Xiao, B.~van Oudheusden, and H.~Bijl, ``A dynamic mesh strategy applied to the simulation of flapping wings,'' \emph{International Journal for Numerical Methods in Engineering}, vol. 106, no.~8, pp. 664--680, 2016.

\bibitem{Delaisse2021}
N.~Delaiss{\'e}, T.~Demeester, D.~Fauconnier, and J.~Degroote, ``Comparison of different quasi-newton techniques for coupling of black box solvers,'' in \emph{14th World Congress on Computational Mechanics/8th European Congress on Computational Methods in Applied Sciences and Engineering}, 2021.

\bibitem{Yeo2010}
K.~Yeo, S.~Ang, and C.~Shu, ``Simulation of fish swimming and manoeuvring by an svd-gfd method on a hybrid meshfree-cartesian grid,'' \emph{Computers \& Fluids}, vol.~39, no.~3, pp. 403--430, 2010.

\bibitem{Zhang2019}
C.~Zhang, T.~L. Hedrick, and R.~Mittal, ``An integrated study of the aeromechanics of hovering flight in perturbed flows,'' \emph{AIAA Journal}, vol.~57, no.~9, pp. 3753--3764, 2019.

\bibitem{Wu2009}
J.~H. Wu, Y.~L. Zhang, and M.~Sun, ``Hovering of model insects: simulation by coupling equations of motion with {N}avier--{S}tokes equations,'' \emph{Journal of Experimental Biology}, vol. 212, no.~20, pp. 3313--3329, 2009.

\bibitem{Liu2010}
H.~Liu, T.~Nakata, N.~Gao, M.~Maeda, H.~Aono, and W.~Shyy, ``Micro air vehicle-motivated computational biomechanics in bio-flights: aerodynamics, flight dynamics and maneuvering stability,'' \emph{Acta Mechanica Sinica}, vol.~26, no.~6, pp. 863--879, 2010.

\bibitem{wu2014}
D.~Wu, K.~Yeo, and T.~Lim, ``A numerical study on the free hovering flight of a model insect at low reynolds number,'' \emph{Computers \& Fluids}, vol. 103, pp. 234--261, 2014.

\bibitem{Yao2018}
Y.~Yao and K.~Yeo, ``Longitudinal free flight of a model insect flyer at low reynolds number,'' \emph{Computers \& Fluids}, vol. 162, pp. 72--90, 2018.

\bibitem{Yao2019v2}
J.~Yao and K.~Yeo, ``Free hovering of hummingbird hawkmoth and effects of wing mass and wing elevation,'' \emph{Computers \& Fluids}, vol. 186, pp. 99--127, 2019.

\bibitem{Taha2012}
H.~E. Taha, M.~R. Hajj, and A.~H. Nayfeh, ``Flight dynamics and control of flapping-wing {MAV}s: a review,'' \emph{Nonlinear Dynamics}, vol.~70, pp. 907--939, 2012.

\bibitem{Orlowski2011}
C.~T. Orlowski and A.~R. Girard, ``Modeling and simulation of nonlinear dynamics of flapping wing micro air vehicles,'' \emph{AIAA journal}, vol.~49, no.~5, pp. 969--981, 2011.

\bibitem{Etkin1995}
B.~Etkin and L.~D. Reid, \emph{Dynamics of flight: stability and control}.\hskip 1em plus 0.5em minus 0.4em\relax John Wiley \& Sons, 1995.

\bibitem{Vanella2010}
M.~Vanella, ``{A fluid structure interaction strategy with application to low Reynolds number flapping flight},'' Ph.D. dissertation, University of Maryland, 2010.

\bibitem{Bakhshaei2021}
K.~Bakhshaei, H.~MoradiMaryamnegari, S.~SalavatiDezfouli, A.~M. Khoshnood, and M.~Fathali, ``Multi-physics simulation of an insect with flapping wings,'' \emph{Proceedings of the Institution of Mechanical Engineers, Part G: Journal of Aerospace Engineering}, vol. 235, no.~10, pp. 1318--1339, 2021.

\bibitem{Xue2023}
Y.~Xue, X.~Cai, R.~Xu, and H.~Liu, ``Wing kinematics-based flight control strategy in insect-inspired flight systems: Deep reinforcement learning gives solutions and inspires controller design in flapping mavs,'' \emph{Biomimetics}, vol.~8, no.~3, p. 295, 2023.

\bibitem{Biswal2019}
S.~Biswal, M.~Mignolet, and A.~A. Rodriguez, ``Modeling and control of flapping wing micro aerial vehicles,'' \emph{Bioinspiration \& biomimetics}, vol.~14, no.~2, p. 026004, 2019.

\bibitem{Chen2024}
H.~Chen, T.~A. Medina, and J.~L. Cercos-Pita, ``{CFD} simulation of multiple moored floating structures using openfoam: An open-access mooring restraints library,'' \emph{Ocean Engineering}, vol. 303, p. 117697, 2024.

\bibitem{Karola2024}
A.~Karola, S.~Tavakoli, T.~Mikkola, J.~Matusiak, and S.~Hirdaris, ``The influence of wave modelling on the motions of floating bodies,'' \emph{Ocean Engineering}, vol. 306, p. 118067, 2024.

\bibitem{Siciliano2008}
B.~Siciliano, O.~Khatib, and T.~Kr{\"o}ger, \emph{Springer handbook of robotics}.\hskip 1em plus 0.5em minus 0.4em\relax Springer, 2008, vol. 200.

\bibitem{Docquier2013}
N.~Docquier, A.~Poncelet, and P.~Fisette, ``Robotran: a powerful symbolic gnerator of multibody models,'' \emph{Mechanical Sciences}, vol.~4, no.~1, pp. 199--219, 2013.

\bibitem{Hu2005}
W.~Hu, D.~W. Marhefka, and D.~E. Orin, ``Hybrid kinematic and dynamic simulation of running machines,'' \emph{IEEE transactions on robotics}, vol.~21, no.~3, pp. 490--497, 2005.

\bibitem{Featherstone2014}
R.~Featherstone, \emph{Rigid body dynamics algorithms}.\hskip 1em plus 0.5em minus 0.4em\relax Springer, 2014.

\bibitem{Hadzic2006}
H.~Hadzic, ``Development and application of finite volume method for the computation of flows around moving bodies on unstructured, overlapping grids,'' Ph.D. dissertation, 2006.

\bibitem{Ferziger2002}
J.~H. Ferziger and M.~Peri{\'c}, \emph{Computational methods for fluid dynamics}.\hskip 1em plus 0.5em minus 0.4em\relax Springer, 2002.

\bibitem{Flores2015}
P.~Flores, \emph{Concepts and formulations for spatial multibody dynamics}.\hskip 1em plus 0.5em minus 0.4em\relax Springer, 2015.

\bibitem{rbd}
``{Rigid body dynamics library},'' \url{https://www.openfoam.com/documentation/guides/latest/api/group__grpRigidBodyDynamics.html}, [Online; accessed 15-June-2024].

\bibitem{Mirtich1996}
B.~V. Mirtich, \emph{Impulse-based dynamic simulation of rigid body systems}.\hskip 1em plus 0.5em minus 0.4em\relax University of California, Berkeley, 1996.

\bibitem{Newmark1959}
N.~M. Newmark, ``A method of computation for structural dynamics,'' \emph{Journal of the engineering mechanics division}, vol.~85, no.~3, pp. 67--94, 1959.

\bibitem{Dullweber1997}
A.~Dullweber, B.~Leimkuhler, and R.~McLachlan, ``Symplectic splitting methods for rigid body molecular dynamics,'' \emph{The Journal of chemical physics}, vol. 107, no.~15, pp. 5840--5851, 1997.

\bibitem{Crank1947}
J.~Crank and P.~Nicolson, ``A practical method for numerical evaluation of solutions of partial differential equations of the heat-conduction type,'' in \emph{Mathematical proceedings of the Cambridge philosophical society}, vol.~43, no.~1.\hskip 1em plus 0.5em minus 0.4em\relax Cambridge University Press, 1947, pp. 50--67.

\bibitem{ofWiki}
``{OverPimpleDyMFoam},'' \url{https://openfoamwiki.net/index.php/OverPimpleDyMFoam}, 2022, [Online; accessed 1-May-2024].

\bibitem{altshuler2012}
D.~L. Altshuler, E.~M. Quicaz{\'a}n-Rubio, P.~S. Segre, and K.~M. Middleton, ``Wingbeat kinematics and motor control of yaw turns in anna's hummingbirds (calypte anna),'' \emph{Journal of experimental biology}, vol. 215, no.~23, pp. 4070--4084, 2012.

\bibitem{poletti2024}
R.~Poletti, A.~Calado, L.~K. Koloszar, J.~Degroote, and M.~A. Mendez, ``On the unsteady aerodynamics of flapping wings under dynamic hovering kinematics,'' \emph{Physics of Fluids}, vol.~36, no.~8, 2024.

\bibitem{Lee2016}
Y.~Lee, K.-B. Lua, T.~Lim, and K.~Yeo, ``A quasi-steady aerodynamic model for flapping flight with improved adaptability,'' \emph{Bioinspiration \& biomimetics}, vol.~11, no.~3, p. 036005, 2016.

\bibitem{Chin2016}
D.~D. Chin and D.~Lentink, ``Flapping wing aerodynamics: from insects to vertebrates,'' \emph{Journal of Experimental Biology}, vol. 219, no.~7, pp. 920--932, 2016.

\bibitem{Berman2007}
G.~J. Berman and Z.~J. Wang, ``Energy-minimizing kinematics in hovering insect flight,'' \emph{Journal of fluid mechanics}, vol. 582, pp. 153--168, 2007.

\bibitem{Yan2015}
Z.~Yan, H.~E. Taha, and M.~R. Hajj, ``Effects of aerodynamic modeling on the optimal wing kinematics for hovering mavs,'' \emph{Aerospace Science and Technology}, vol.~45, pp. 39--49, 2015.

\bibitem{Bayiz2018}
Y.~Bayiz, M.~Ghanaatpishe, H.~Fathy, and B.~Cheng, ``Hovering efficiency comparison of rotary and flapping flight for rigid rectangular wings via dimensionless multi-objective optimization,'' \emph{Bioinspiration \& biomimetics}, vol.~13, no.~4, p. 046002, 2018.

\bibitem{Bhat2020}
S.~S. Bhat, J.~Zhao, J.~Sheridan, K.~Hourigan, and M.~C. Thompson, ``Effects of flapping-motion profiles on insect-wing aerodynamics,'' \emph{Journal of Fluid Mechanics}, vol. 884, p.~A8, 2020.

\bibitem{Ellington1984_III}
C.~Ellington, ``{The aerodynamics of hovering insect flight. III. Kinematics},'' \emph{Philosophical Transactions of the Royal Society of London. B, Biological Sciences}, vol. 305, no. 1122, pp. 41--78, 1984.

\bibitem{Kruyt2014}
J.~W. Kruyt, E.~M. Quicaz{\'a}n-Rubio, G.~F. Van~Heijst, D.~L. Altshuler, and D.~Lentink, ``Hummingbird wing efficacy depends on aspect ratio and compares with helicopter rotors,'' \emph{Journal of the {R}oyal {S}ociety {I}nterface}, vol.~11, no.~99, p. 20140585, 2014.

\bibitem{Bos2013}
F.~M. Bos, B.~W. van Oudheusden, and H.~Bijl, ``Wing performance and 3-d vortical structure formation in flapping flight,'' \emph{Journal of Fluids and Structures}, vol.~42, pp. 130--151, 2013.

\bibitem{Nakata2015}
T.~Nakata, H.~Liu, and R.~J. Bomphrey, ``A {CFD}-informed quasi-steady model of flapping-wing aerodynamics,'' \emph{Journal of Fluid Mechanics}, vol. 783, pp. 323--343, 2015.

\bibitem{Badrya2016}
C.~Badrya, ``{CFD/Quasi-Steady coupled trim analysis of diptera-type flapping wing MAV in steady flight},'' Ph.D. dissertation, University of Maryland, College Park, 2016.

\bibitem{Calado2023}
A.~Calado, R.~Poletti, L.~K. Koloszar, and M.~A. Mendez, ``A robust data-driven model for flapping aerodynamics under different hovering kinematics,'' \emph{Physics of Fluids}, vol.~35, no.~4, 2023.

\bibitem{Alletto2022}
M.~Alletto, ``{Comparison of overset mesh with morphing mesh: flow over a forced oscillating and freely oscillating 2D cylinder},'' \emph{OpenFOAM{\textregistered} Journal}, vol.~2, pp. 13--30, 2022.

\bibitem{Coe2023}
M.~Coe and S.~Gutschmidt, ``Ika-flow: A flexible body overset mesh implementation for fish swimming,'' \emph{OpenFOAM{\textregistered} Journal-V3. 4-Closed}, vol.~3, pp. 75--119, 2023.

\bibitem{Ahmed2024}
K.~Ahmed, A.~A. Hamada, L.~Chatellier, and M.~Furth, ``A modified overset method in openfoam for simultaneous motion and deformation: A case study of a flexible flapping foil,'' \emph{OpenFOAM{\textregistered} Journal}, vol.~4, pp. 41--61, 2024.

\bibitem{Stachowiak2006}
T.~Stachowiak and T.~Okada, ``A numerical analysis of chaos in the double pendulum,'' \emph{Chaos, Solitons \& Fractals}, vol.~29, no.~2, pp. 417--422, 2006.

\bibitem{Calvao2015}
A.~Calv{\~a}o and T.~Penna, ``The double pendulum: a numerical study,'' \emph{European Journal of Physics}, vol.~36, no.~4, p. 045018, 2015.

\bibitem{Herho2024}
S.~Herho, F.~Fajary, K.~Herho, I.~Anwar, R.~Suwarman, and D.~E. Irawan, ``Reappraising double pendulum dynamics across multiple computational platforms,'' 2024.

\bibitem{Schubert2017}
S.~Schubert and T.~Rung, ``Challenges and applications of overset grid coupling strategies,'' \emph{PAMM}, vol.~17, no.~1, pp. 139--142, 2017.

\bibitem{Sane2002}
S.~P. Sane and M.~H. Dickinson, ``The aerodynamic effects of wing rotation and a revised quasi-steady model of flapping flight,'' \emph{Journal of experimental biology}, vol. 205, no.~8, pp. 1087--1096, 2002.

\bibitem{Schena2023}
L.~Schena, P.~Marques, R.~Poletti, S.~Ahizi, J.~V.~d. Berghe, and M.~A. Mendez, ``Reinforcement twinning: from digital twins to model-based reinforcement learning,'' \emph{arXiv preprint arXiv:2311.03628}, 2023.

\bibitem{Li2022}
H.~Li and M.~R. Nabawy, ``{Capturing wake capture: a 2D numerical investigation into wing--wake interaction aerodynamics},'' \emph{Bioinspiration \& Biomimetics}, vol.~17, no.~6, p. 066015, 2022.

\bibitem{Van2022}
W.~G. Van~Veen, J.~L. Van~Leeuwen, B.~W. Van~Oudheusden, and F.~T. Muijres, ``The unsteady aerodynamics of insect wings with rotational stroke accelerations, a systematic numerical study,'' \emph{Journal of Fluid Mechanics}, vol. 936, p.~A3, 2022.

\bibitem{Windt2018}
C.~Windt, J.~Davidson, B.~Akram, and J.~V. Ringwood, ``Performance assessment of the overset grid method for numerical wave tank experiments in the openfoam environment,'' in \emph{International Conference on Offshore Mechanics and Arctic Engineering}, vol. 51319.\hskip 1em plus 0.5em minus 0.4em\relax American Society of Mechanical Engineers, 2018, p. V010T09A006.

\bibitem{oversetImpl}
``{Overset implementation},'' \url{https://github.com/louisgag/openFoam-Overset-SpeedUp}, 2022, [Online; accessed 15-June-2024].

\bibitem{Jeong1995}
J.~Jeong and F.~Hussain, ``On the identification of a vortex,'' \emph{Journal of Fluid Mechanics}, vol. 285, pp. 69--94, 02 1995.

\bibitem{kratos}
``{kratos},'' \url{https://github.com/KratosMultiphysics/Kratos}, [Online; accessed 18-June-2024].

\bibitem{coconut}
``{coconut},'' \url{https://pyfsi.github.io/coconut/}, [Online; accessed 18-June-2024].

\end{thebibliography}




\end{document}